# Phosphine Gas in the Cloud Decks of Venus

Jane S. Greaves[1*†], Anita M. S. Richards[2], William Bains[3], Paul B. Rimmer[4,5,6], Hideo Sagawa[7], David L. Clements[8], Sara Seager[3‡§], Janusz J. Petkowski[3], Clara Sousa-Silva[3], Sukrit Ranjan[3¶], Emily Drabek-Maunder[1,9], Helen J. Fraser[10], Annabel Cartwright[1], Ingo Mueller-Wodarg[8], Zhuchang Zhan[3], Per Friberg[11], Iain Coulson[11], E'lisa Lee[11], Jim Hoge[11].

[1]School of Physics & Astronomy, Cardiff University, 4 The Parade, Cardiff CF24 3AA, UK.
[2]Jodrell Bank Centre for Astrophysics, Department of Physics and Astronomy, The University of Manchester, Alan Turing Building, Oxford Road, Manchester, M13 9PL, UK.
[3]Department of Earth, Atmospheric, and Planetary Sciences, Massachusetts Institute of Technology, 77 Mass. Ave., Cambridge, MA, 02139, USA.
[4]Department of Earth Sciences, University of Cambridge, Downing Street, Cambridge CB2 3EQ, UK.
[5]Cavendish Astrophysics, University of Cambridge, JJ Thomson Avenue, Cambridge CB3 0HE, UK.
[6]MRC Laboratory of Molecular Biology, Francis Crick Ave, Trumpington, Cambridge CB2 0QH, UK.
[7]Department of Astrophysics and Atmospheric Science, Kyoto Sangyo University, Kyoto 603-8555, Japan.
[8]Department of Physics, Imperial College London, South Kensington Campus, London SW7 2AZ, UK.
[9]Royal Observatory Greenwich, Blackheath Ave, Greenwich, London SE10 8XJ, UK.
[10]School of Physical Sciences, The Open University, Walton Hall, Milton Keynes MK7 6AA, UK.
[11]East Asian Observatory, 660 N. A'ohoku Place, University Park, Hilo, HI 96720, USA.
*Correspondence to: greavesj1@cardiff.ac.uk.
†Visitor at Institute of Astronomy, Cambridge University, Madingley Rd, Cambridge CB3 0HA, UK.
‡Department of Physics and Kavli Institute for Astrophysics and Space Research, Massachusetts Institute of Technology, 77 Mass. Ave., Cambridge, MA, 02139, USA.
§Dept. of Aeronautics and Astronautics, Massachusetts Institute of Technology, 77 Mass. Ave., Cambridge, MA, 02139, USA.
¶SCOL Postdoctoral Fellow

**Measurements of trace-gases in planetary atmospheres help us explore chemical conditions different to those on Earth. Our nearest neighbor, Venus, has cloud decks that are temperate but hyper-acidic. We report the apparent presence of phosphine ($PH_3$) gas in Venus' atmosphere, where any phosphorus should be in oxidized forms. Single-line millimeter-waveband spectral detections (quality up to ~15$\sigma$) from the JCMT and ALMA telescopes have no other plausible identification. Atmospheric $PH_3$ at ~20 parts-per-billion abundance is inferred. The presence of phosphine is unexplained after exhaustive study of steady-state chemistry and photochemical pathways, with no currently-known abiotic production routes in Venus' atmosphere, clouds, surface and subsurface, or from lightning, volcanic or meteoritic delivery. Phosphine could originate from unknown photochemistry or geochemistry, or, by analogy with biological production of phosphine on Earth, from the presence of life. Other $PH_3$ spectral features should be sought, while in-situ cloud/surface sampling could examine sources of this gas.**

Studying rocky-planet atmospheres gives clues to how they interact with surfaces and subsurfaces, and whether any non-equilibrium compounds could reflect the presence of life. Characterizing extrasolar-planet atmospheres is extremely challenging, especially for rare compounds[1]. The solar system thus offers important testbeds for exploring planetary geology, climate and habitability, via both *in-situ* sampling and remote-monitoring. Proximity makes signals of trace gases much stronger than those from extrasolar planets, but issues remain in interpretation.

Thus far, solar system exploration has found compounds of interest, but often in locations where the gas-sources are inaccessible, such as the Martian sub-surface[2] and water-reservoirs inside icy moons[3,4]. Water, simple organics and larger unidentified carbon-bearing species[5-7] are known. However, geochemical sources for carbon-compounds may exist[8], and temporal/spatial anomalies can be hard to reconcile, e.g. for Martian methane sampled by rovers and observed from orbit[9].

An ideal biosignature-gas would be unambiguous. Living organisms should be its sole source, and it should have intrinsically-strong, precisely-characterized spectral transitions unblended with contaminant-lines – criteria not usually all achievable. It was recently proposed that any phosphine detected in a rocky-planet's atmosphere is a promising sign of life[10]. Trace $PH_3$ in the Earth's atmosphere (parts-per-trillion abundance globally[11]) is uniquely associated with anthropogenic activity or microbial presence – life produces this highly-reducing gas even in an overall oxidizing environment. Phosphine is found elsewhere in the solar system only in the reducing atmospheres of giant planets[12,13], where it is produced in deep atmospheric layers at high temperatures and pressures, and dredged upwards by convection[14,15]. Solid surfaces of rocky planets present a barrier to their interiors, and phosphine would be rapidly destroyed in their highly-oxidized crusts and atmospheres. Thus $PH_3$ meets most criteria for a biosignature-gas search, but is challenging as many of its spectral features are strongly absorbed by the Earth's atmosphere.

Motivated by these considerations, we exploited the $PH_3$ 1-0 millimeter-waveband rotational-transition that could absorb against optically-thick layers of Venus' atmosphere. Long-standing speculations consider aerial biospheres in the high-altitude cloud decks[16,17], where conditions have some similarity to ecosystems making phosphine on Earth[18]. We exploited good instrument sensitivity, 25 years after the first millimeter-waveband exploration of solar-system $PH_3$ (in Saturn's atmosphere[19]). We proposed a 'toy-model' experiment that could set abundance-limits of order parts-per-billion on Venus, comparable to phosphine production of some anaerobic Earth ecosystems[10]. The aim was a benchmark for future developments, but unexpectedly, our initial observations suggested a detectable amount of Venusian phosphine was present.

We present next the discovery data, confirmation (and preliminary mapping) by follow-up observations, and rule out line-contamination. We then address whether gas reactions, photo/geo-chemical reactions or exogenous non-equilibrium input could plausibly produce $PH_3$ on Venus.

## Results

The $PH_3$ 1-0 rotational transition at 1.123 mm wavelength was initially sought with the James Clerk Maxwell Telescope (JCMT), in observations of Venus over 5 mornings in June 2017. The single-point spectra cover the whole planet (limb down-weighted by ~50% within the telescope beam). Absorption lines from the cloud decks were sought against the quasi-continuum created by overlapping broad emission features from the deeper, opaque atmosphere.

The main limitation at small line-to-continuum ratio (hereafter, l:c) was spectral 'ripple', from artefacts such as signal reflections. We identified three issues (see Methods: JCMT data reduction), with the most-problematic being high-frequency ripple drifting within observations in a manner hard to remove even in Fourier space (Supplementary Figure 1). We thus followed an approach standardised over several decades[20], fitting amplitude-versus-wavelength polynomials to the ripples (in 140 spectra). The passband was truncated to 100 km/s to avoid using high polynomial-orders. (Order is based on the number N of 'bumps' in the ripple-pattern; fitting is optimal with order N+1 and negligibly improved at increased order. A wider band includes more 'bumps', increasing N. For minimum freedom, a linear fit can be employed immediately around the line-candidate, ignoring the remaining passband – see Table 1 for resulting systematic differences.) We explored a range of solutions with the spectra flattened outside a velocity-interval within which absorption is allowed. (The polynomial must be interpolated across an interval, since if fitted to the complete band it will always remove a line-candidate, given freedom to increase order.) These interpolation-intervals ranged from very narrow, preserving only the line core (predicted by our radiative-transfer models, Figure 1), up to a Fourier-defined limit above which negative-sign artefacts can mimic an absorption line. Details are in Methods, with the reduction script in Supplementary Software 1. The spectra were also reduced completely independently by a second team member, via a minimal-processing method that collapses the data-stack down the time-axis and fits a one-step low-order polynomial; this gave a similar output-spectrum but with lower signal-to-noise.

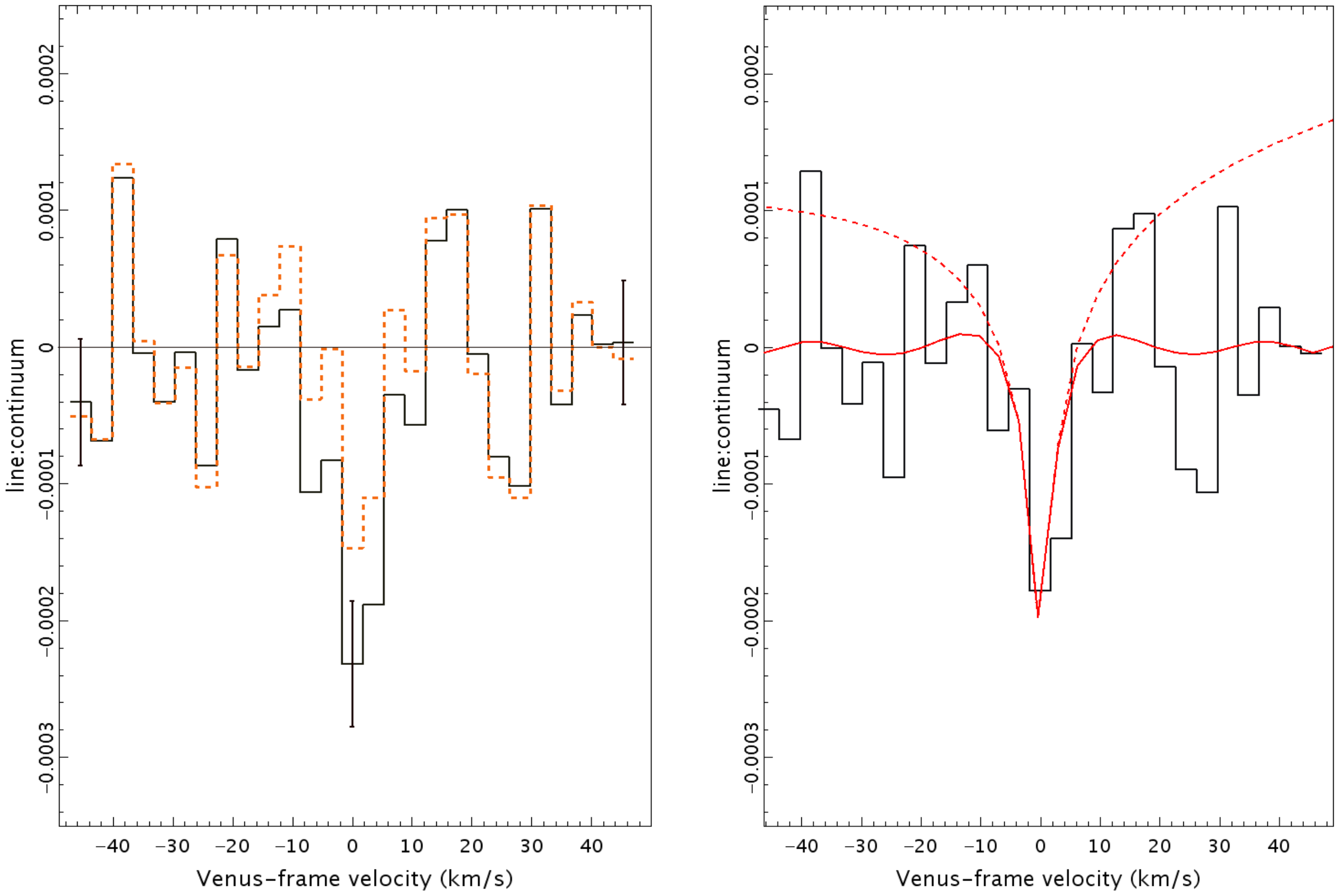

**Fig. 1.** Panels show spectra of $PH_3$ 1-0 in Venus' atmosphere as observed with the JCMT. Axes are line-to-continuum ratio against Doppler-shifted velocity referenced to the phosphine wavelength. Left: the least and most conservative solutions after fitting and removing spectral ripple (see SI: JCMT data analysis), with the residual line present inside velocity-ranges of |v| = 8 km/s (solid, black) and |v| = 2 km/s (dashed, orange). The data have been binned into histograms (bars denoting averages) on the x-axis for clarity; representative 1σ error bars are $0.46 \cdot 10^{-4}$ in l:c per 3.5 km/s spectral bin. Error bars indicate the dispersion within each channel from 140 co-added input spectra; channel-to-channel dispersion is higher by ~40%, attributable to residual ripple, and contributing to the range of signal-to-noise (Table 1). Right: the adopted mid-range solution with |v| = 5 km/s (histogram), overlaid with our model for 20 ppb abundance-by-volume. The solid red curve shows this model after processing with the same spectral fitting as used for the data. The line wings and continuum slope have thus been removed from the original model (upper dashed red curve). As the spectral fitting forces the line wings towards zero, only the range ±10 km/s around Venus' velocity was used in line characterisation (Table 1).

In our co-added spectrum (Fig. 1), we saw candidate $PH_3$ 1-0 absorption, with signal-to-noise varying over ~3-7, depending on the velocity-interval selection. The feature is consistent with Venus' velocity, but is not precisely characterized (Table 1). This potentially allows for the feature to be a weak residual artefact, or a transition of another molecule at a nearby wavelength.

We thus sought confirmation of the same transition, with independent technology and improved signal-to-noise, using the Atacama Large Millimetre/submillimetre Array (ALMA) in March 2019. In principle, ALMA's arcsecond-scale resolution would allow detailed mapping of the planet's atmosphere. In practice, interferometric response to a large bright planet produced artefactual spectral ripples varying from baseline to baseline (and not eliminated by bandpass calibration). This systematic was greatly reduced, prior to imaging, by excluding all telescope-to-telescope baselines < 33 m in length. This was necessary for dynamic range and was the only significant departure from the standard ALMA 'QA2' approach[21] to data reduction (see Methods: ALMA data reduction; Supplementary Figures 2-4; Supplementary Software 2-4). While bandpass calibration using Jupiter's moon Callisto was not fully sufficient, the dynamic range achieved was still substantially higher than ALMA's specification (~$10^{-3}$ in l:c, without the techniques we used to reduce systematics, and which we verified did not produce spurious features). To eliminate residual ripple from the extracted spectra, we tested polynomial-fitting strategies with orders ranging from 12 (optimal for an 80 km/s passband, Figure), down to 1 (fitting only around the line-candidate). The resulting systematic uncertainties are summarised in Table 1.

We also checked for robustness by searching simultaneously for deuterated water (HDO) known to be present on Venus. The HDO $2_{2,0}$-$3_{1,3}$ line at 1.126 mm wavelength was detected (Supplementary Figure 5: preliminary output from manual 'QA2' scripts), with a line profile well-fitted by our radiative transfer model, and a Venus-normal water abundance (see Methods: ALMA data reduction). Simultaneous wider-bandpass settings also allowed us to set upper limits on other chemical species – transitions here could be a check on possible contaminants, i.e. constrain transitions close in wavelength to the line we identify as $PH_3$ 1-0. The wide-bandpass tuning centred on this $PH_3$ transition provided a further reproducibility check. These data have significantly greater problems with spectral ripple than in the narrow-bandpass settings, but the phosphine line was recovered (Supplementary Figure 6).

The effect of removal of short-baseline ALMA data is that line-signals from areas smooth on scales > 4 arcseconds are substantially diluted. Thus our l:c correspond to lower limits of $PH_3$ abundance (but detection-significance is not affected; these values are as stated in Table 1). Further, the steep flux density gradients at the limb resulted in more flux being recovered here. To ensure that results are robust, we did not attempt to interpret any absorption spectra over arcsec-scales (see Supplementary Figure 3). To mitigate for the bias in better-sampling the limb, the spectra in Figure 2 are all averages from 'side-to-side' strips across the planet.

The ALMA data confirm the detection of absorption at the $PH_3$ 1-0 wavelength. All line-centroid velocities are consistent with Venus' velocity within -0.2 to +0.7 km/s (around 10% of the line width), with best measurement-precision at ±0.3 km/s and systematics of ~0.1-0.7 km/s (Table 1). For this degree of coincidence of apparent velocity, any contaminating transition from another chemical species would have to coincide in rest-wavelength with $PH_3$ 1-0 within $\sim 10^{-6}$.

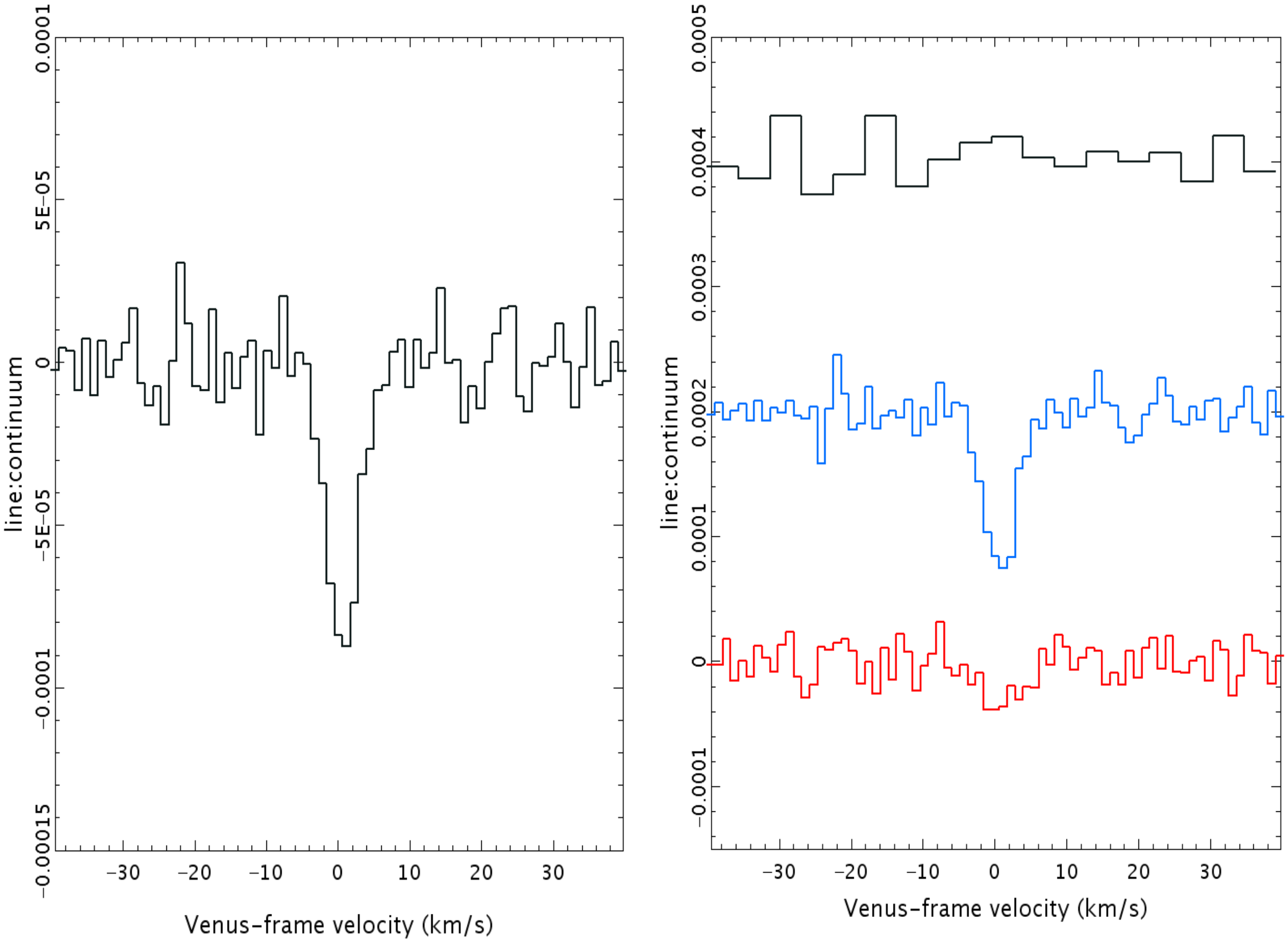

**Fig. 2.** Spectra of Venus obtained with ALMA. Left panel shows the $PH_3$ 1-0 spectrum of the whole planet, with 1σ errors (here channel-to-channel) of 0.11 $10^{-4}$ per 1.1 km/s spectral bin. Right panel shows spectra of the polar (histogram in black), mid-latitude (in blue) and equatorial (in red) zones, as defined in Table 1. Spectra have been offset vertically for clarity, and the polar spectrum was binned in velocity to obtain a deeper upper limit. Line wings are forced towards zero outside |v| = 5 km/s in these spectra, and only this range was used in characterization (Table 1; Methods: ALMA data reduction).

| facility (epoch) | area of planet | line-to-continuum ratio ($10^{-4}$) | centroid (km/s) | FWHM (km/s) | signal-to-noise ratio | Notes |
|---|---|---|---|---|---|---|
| JCMT (June, 2017) | whole planet | -2.5 ± 0.8 (-2.2, -3.1) | -0.2 ± 1.1 (-0.3 ± 1.2, -0.3 ± 0.9) | 3.6 ± 1.2 (2.8 ± 1.0, 8.2 ± 2.3) | 4.3 (3.0, 6.7) | \|v\| = 5 km/s (\|v\| = 2,8 km/s for systematics) |
| ALMA (March, 2019) | whole planet | -0.87 ± 0.11 | +0.7 ± 0.3 (+0.3 ± 0.3) | 4.1 ± 0.5 | 13.3 | \|v\| = 5 km/s (linear fit for systematic) |
| | equator (15°S-15°N) | -0.39 ± 0.14 | +0.7 ± 0.9 (-0.0 ± 0.4) | 4.8 ± 1.8 | 5.0 | as for whole planet |
| | mid-latitude (15-60°S + 15-60°N) | -1.26 ± 0.14 | +0.7 ± 0.3 (+0.4 ± 0.3) | 4.1 ± 0.6 | 14.5 | as for whole planet |
| | polar (60-90°S + 60-90°N) | (3σ: -0.29) | --- | --- | --- | limit for 10 km/s bins |

**Table 1.** Properties of the absorption line for regions of Venus' atmosphere. Measurement errors are 1σ, and systematic errors are differences of the means and the mean values in brackets, the latter being obtained with the data-processing modifications stated in the 'Notes' column. Line-to-continuum ratios are measured at line-minimum, for 1.1 km/s spectral bins that are in common to both datasets. Centroid velocities are referenced to the $PH_3$ 1-0 line-identification. Lines were fitted with Lorentzian profiles over ±10 km/s to estimate full-width half-minima (FWHM). For JCMT, intensity-weighted velocity centroids and line-integrated signal-to-noise (based on per-channel errors) were calculated over ±10 km/s velocity-ranges. For ALMA, calculation ranges were restricted to ±5 km/s because of complexity of spectral ripple (see Supplementary Figure 4), and centroids in brackets are for comparison, from a simplified linear fit immediately adjacent to the absorption. In all other cases, results are from spectra in Figures 1 and 2, after the removal of polynomial baselines of order 8 (JCMT) and 12 (ALMA). We verified that high-order fitting does not produce artefact lines at arbitrary positions in the passband (Supplementary Figure 4).

The data above represent the candidate discovery of phosphine on Venus. Because of the very high l:c sensitivity required, we tested robustness through several routes. In particular, we analysed data from both facilities by a range of methods and estimated systematic uncertainties.

The JCMT and ALMA whole-planet spectra agree in line-velocity and width, and are consistent in line-depth after taking into account ALMA's spatial filtering (hence, no temporal-variation in $PH_3$ abundance needs to be invoked over 2017-2019). We considered ALMA's maximum line-loss, in the case of a phosphine distribution as uniform as the almost-smooth continuum (Supplementary Figure 2). Comparing the ALMA continuum signals with/without baselines of <33 m in the data reduction, we found filtering-losses varying from a net 60% in our polar regions to 92% for our equatorial band. Correcting the whole-planet line-signal by this method, l:c could rise from $-0.9 \cdot 10^{-4}$ to $-4.9 \cdot 10^{-4}$, values bracketing $-2.5 \cdot 10^{-4}$ from the JCMT. Hence, the ALMA and JCMT lines differ by factors of at most 2-3, with agreement possible if the phosphine is distributed on intermediate scales (between highly-uniform and small patches).

Finally, for robustness, we considered the possibility of a 'double-false-positive', where a negative-dip occurs in both datasets near the Venusian velocity. Comparing the data *before* the final processing-step of polynomial-fitting take place, Figure 3 shows that no other coincidences of absorption-line-like features occur in the JCMT and ALMA spectra.

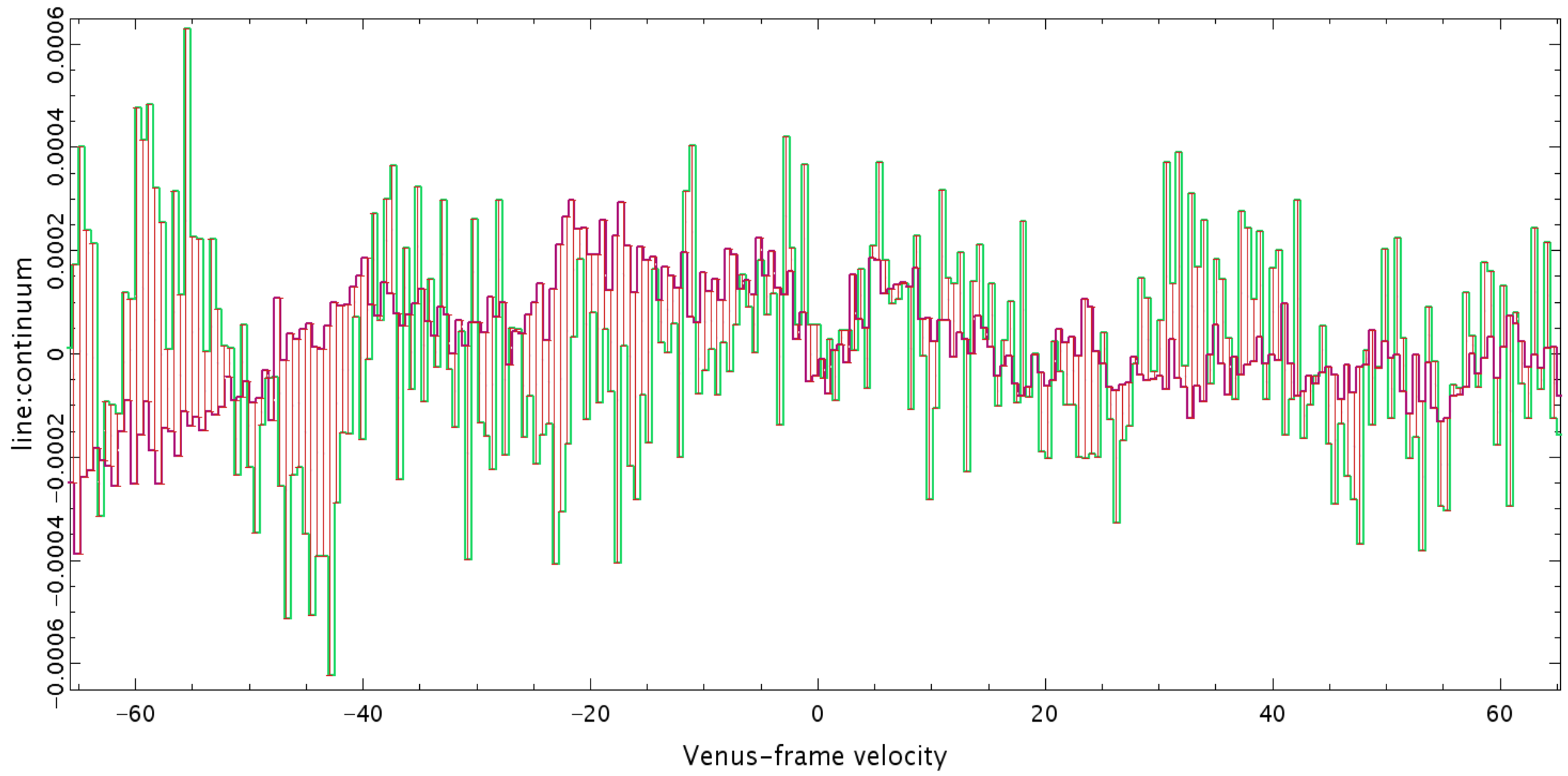

**Fig. 3.** JCMT and ALMA whole-planet spectra (green and purple histograms, respectively), across the full passband in common. These are the co-added spectra *before* the removal of a final polynomial baseline. The ALMA spectrum has been scaled up by a factor of 3, the estimated loss for spatial filtering (compare the first two l:c entries in Table 1). Vertical red bars connect the JCMT and ALMA data (their spectral bin-centres agree in velocity within ±0.2 km/s). A line feature is considered to be real where this dispersion (red bar) is low, and only the candidate phosphine feature around v = 0 km/s meets this criterion. Other candidate 'dips' across the band have high dispersion (as they occur only in one dataset), or cover only a few contiguous bins (much less than the line-width expected for Venusian upper-atmosphere absorption).

Next, we examined whether transitions from gases other than $PH_3$ might absorb at nearby wavelengths. The only plausible candidate (Supplementary Table 1) is an $SO_2$ transition offset by +1.3 km/s in the reference frame of $PH_3$ 1-0. This is expected to produce a weak line in the cloud decks, with its lower quantum-level at energy > 600 K not being highly-populated in < 300 K gas. $SO_2$ absorptions from energy-levels at ~100 K have been detected[22], and we searched for one such transition in our simultaneous ALMA wideband-data. We did not detect significant absorption (Figure 4a). Given this observation, our radiative-transfer model predicts what the maximum absorption from the 'contaminant' $SO_2$ line would be, finding a weak l:c, not deeper than $-0.2\cdot10^{-4}$ (Figure 4b). $SO_2$ can contribute a maximum of <10% to the l:c integrated over ±5 km/s, and shift the line-centroid by <0.1 km/s. These results are abundance- and model-independent. The contaminant-$SO_2$-line could only 'mimic' the phosphine feature while the wideband-$SO_2$-line remained undetected if the gas were more than twice as hot as measured in the upper clouds – i.e. at temperatures only found at much lower altitudes than our data probe.

We are unable to find another chemical species (known in current databases[23-26]) besides $PH_3$ that can explain the observed features. We conclude that the candidate detection of phosphine is robust, for four main reasons. Firstly, the absorption has been seen, at comparable line depth, with two independent facilities; secondly, line-measurements are consistent under varied and independent processing methods; thirdly, overlap of spectra from the two facilities shows no other such consistent negative features; and fourthly, there is no other known reasonable candidate-transition for the absorption other than phosphine.

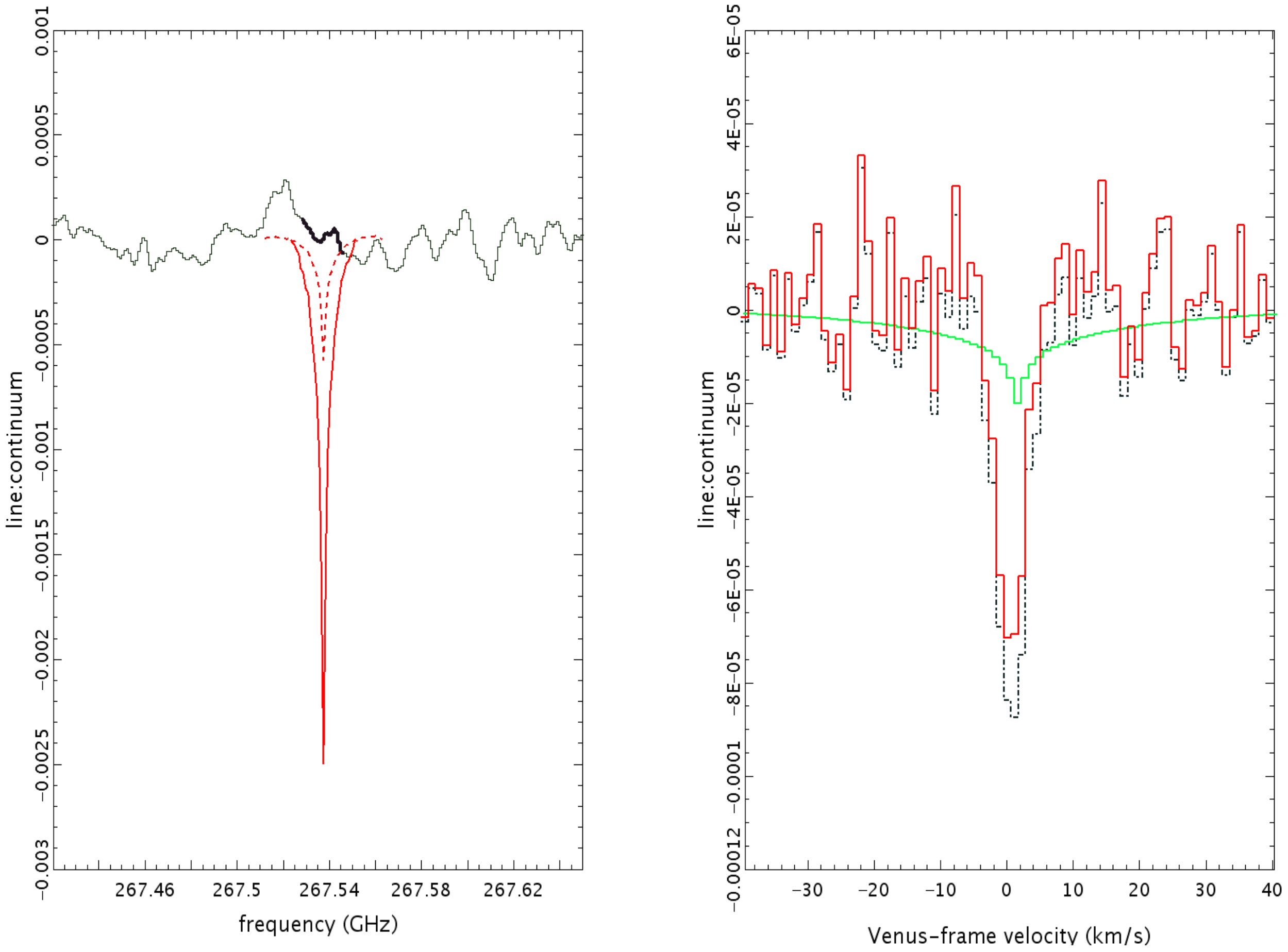

**Fig. 4.** Left panel shows a section of ALMA wideband data (whole planet, after a 3$^{rd}$-order polynomial correcting for broad curvature has been removed), around the $SO_2$ $13_{3,11}$-$13_{2,12}$ rest-frequency (267.53745 GHz; wavelength ≈ 1.121 mm). The thicker histogram over ±10 km/s range illustrates that $SO_2$ absorption is not seen. The red dashed curve is an $SO_2$-10 ppb model, after subtracting a polynomial forcing line wings towards zero outside |v| = ±10 km/s. The 10 ppb model was chosen to reproduce the maximum line depth possible within the data, approximating to the peak-to-peak spectral ripple. The red solid curve is scaled up to show the amplitude this $SO_2$ line would need to have if the line we identify as $PH_3$ 1-0 is instead all attributed to the $SO_2$ $30_{9,21}$-$31_{8,24}$ transition. Right panel re-plots our model for the maximum allowed $SO_2$ $30_{9,21}$-$31_{8,24}$ contribution (as in the red dashed model of the left panel, but without the polynomial subtraction: green histogram). The $PH_3$ whole-planet spectrum (black dot-dashed histogram) is then re-plotted (red solid histogram) after subtraction of this maximized level of $SO_2$ $30_{9,21}$-$31_{8,24}$.

The few-km/s widths of the $PH_3$ spectra are typical of absorptions from the upper atmosphere of Venus[22]. Inversion techniques[27] can convert line-profiles into a vertical molecular-distribution, but this is challenging here due to uncertainties in $PH_3$ line-dilution and pressure-broadening. As the continuum against which we see absorption[28] arises at altitudes ~53-61 km (Supplementary Figure 2), in the middle/upper cloud deck layers[17], the $PH_3$ molecules observed must be at least this high up. Here the clouds are 'temperate', at up to ~30$^{o}$C, and with pressures up to ~0.5 bar[29]. However, phosphine could form at lower (warmer) altitudes and then diffuse upwards.

Phosphine is detected most strongly at mid-latitudes, and is not detected at the poles (Table 1). The equatorial zone appears to absorb more weakly than mid-latitudes, but equatorial and mid-latitude values could agree if corrections are made for spatial filtering. Following the method above (treating gas as if distributed like the continuum), then l:c can be as deep as $-4.6 \cdot 10^{-4}$ for the equator and $-5.8 \cdot 10^{-4}$ for mid-latitudes, in agreement at the $1\sigma$ bounds (both $\pm$ $0.7 \cdot 10^{-4}$). However, for the polar caps, l:c cannot exceed $-0.7 \cdot 10^{-4}$ by this method (as small limb regions are the least-affected by missing short-baseline data). Our latitude ranges were set empirically, to maximise contrasts in l:c, so may not represent physical zones. We were unable to compare bands of longitude (e.g. for any effects of Solar angle), as regions nearer the limb had increasing issues of noise and spectral ripple (Supplementary Figure 3).

The abundance of phosphine in Venus' atmosphere was estimated by comparing a model line to the JCMT spectrum, which has the least signal-losses. The radiative transfer in Venus' atmosphere was calculated using a spherical, multi-layered model, with temperature and pressure profiles from the Venus international reference atmosphere (VIRA). Molecular absorptions are calculated by a line-by-line code, including $CO_2$ continuum-induced opacity. JCMT beam-dilution is included. The abundance calculated is ~20 ppb (Figure 1). The model's major uncertainty is in the $CO_2$ pressure-broadening coefficient, which has not been measured for $PH_3$. We take $PH_3$ 1-0 line broadening coefficients to range from 0.186 $cm^{-1}$/atm, (our theoretical estimate) to 0.286 $cm^{-1}$/atm (the measured value for the $CO_2$ broadening of the $NH_3$ 1-0 line). Ammonia and phosphine share many similarities (see Methods: Abundance retrieval), and can be expected to have comparable broadening properties[30,31]. With this range of coefficients, derived abundances range from ~20 ppb (using our theoretical estimate) up to ~30 ppb (using the proxy $NH_3$-broadening). Additionally, uncertainty in l:c in the JCMT spectrum contributes ~30% (±6 ppb), with additional shifts of -2,+5 ppb possible from systematics (Table 1).

The presence of even a few parts-per-billion of phosphine is completely unexpected for an oxidized atmosphere (where oxygen-containing compounds greatly dominate over hydrogen-containing ones). We review all scenarios that could plausibly create phosphine, given established knowledge of Venus.

The presence of $PH_3$ implies an atmospheric, surface or subsurface source of phosphorus, or delivery from interplanetary space. The only measured values of atmospheric P on Venus come from Vega descent probes[32], which were only sensitive to phosphorus as an element, so its chemical speciation is not known. No P-species have been reported at the planetary surface.

The bulk of any P present in Venus' atmosphere or surface is expected as oxidized forms of phosphorus, e.g. phosphates. Considering such forms, and adopting Vega abundance data (the highest inferred value, most favorable for $PH_3$ production), we calculate whether equilibrium thermodynamics under conditions relevant to the Venusian atmosphere, surface, and subsurface can provide ~10 ppb of $PH_3$. (We adopt a lower-bound adequately fitting the JCMT data, to find the most readily-achievable thermodynamic solution.) We find that $PH_3$ formation is not favored even considering ~75 relevant reactions under thousands of conditions encompassing any likely atmosphere, surface, or subsurface properties (temperatures of 270-1500 K, atmospheric and subsurface pressures of 0.25-10,000 bar, wide range of concentrations of reactants). The free energy of reactions falls short by anywhere from 10 to 400 kJ/mol (see "Potential pathways to phosphine production" in Methods, Supplementary Information, Supplementary Figure 7). In particular, we quantitatively rule out the hydrolysis of geological or meteoritic phosphide as the source of Venusian phosphine. We also rule out the formation of phosphorous acid ($H_3PO_3$).

While phosphorous acid can disproportionate to phosphine on heating, its formation under Venus temperatures and pressures would require quite unrealistic conditions, such as an atmosphere composed almost entirely of hydrogen (for details, see Supplementary Information).

The lifetime of phosphine on Venus is key for understanding production rates that would lead to accumulation of few-ppb concentrations. This lifetime will be much longer than on Earth, whose atmosphere contains substantial molecular oxygen and its photochemically-generated radicals. The lifetime above 80 km on Venus (in the mesosphere[22]) is consistently predicted by models to be <$10^3$ seconds, primarily due to high concentrations of radicals that react with, and destroy, $PH_3$. Near the atmosphere's base, estimated lifetime is ~$10^8$ seconds due to thermal-decomposition (collisional-destruction) mechanisms. Lifetimes are very poorly constrained at intermediate altitudes (<80 km), being dependent on abundances of trace radical species, especially chlorine. These lifetimes are uncertain by orders-of-magnitude, but are substantially longer than the time for $PH_3$ to be mixed from the surface to 80 km (< $10^3$ years). The lifetime of phosphine in the atmosphere is thus no longer than $10^3$ years, either because it is destroyed more quickly or because it is transported to a region where it is rapidly destroyed (see "Photochemical model" in Methods, Supplementary Information, Supplementary Figures 8-9, Supplementary Tables 2-3).

We estimate the out-gassing flux of $PH_3$ needed to maintain ~10 ppb levels, taking the column of phosphine derived from observations and dividing this by the chemical lifetime of phosphine in Venus' atmosphere (Figure 5). The total outgassing-flux necessary to explain ~10 ppb of $PH_3$ is ~$10^6$-$10^7$ molecules cm$^{-2}$ s$^{-1}$ (shorter lifetimes would lead to higher flux requirements). Photochemically-driven reactions in Venus' atmosphere cannot produce phosphine at this rate. To generate $PH_3$ from oxidized P-species, photochemically-generated radicals have to reduce the phosphorus by abstracting oxygen and adding hydrogen – requiring reactions predominantly with H, but also with O and OH radicals. Hydrogen-radicals are rare in Venus' atmosphere because of low concentrations of potential hydrogen-sources (species such as $H_2O$, $H_2S$ that are UV-photolyzed to produce H radicals). We model a network of forward-reactions (i.e. from oxidized P-species to $PH_3$), not only as a conservative maximum-possible production rate for $PH_3$, but also because many of the back-reaction rates are not known. We find the reaction rates of H radicals with oxidized phosphorus species are too slow by factors of $10^4$-$10^6$ under the temperatures and concentrations in the Venusian atmosphere (Figure 5).

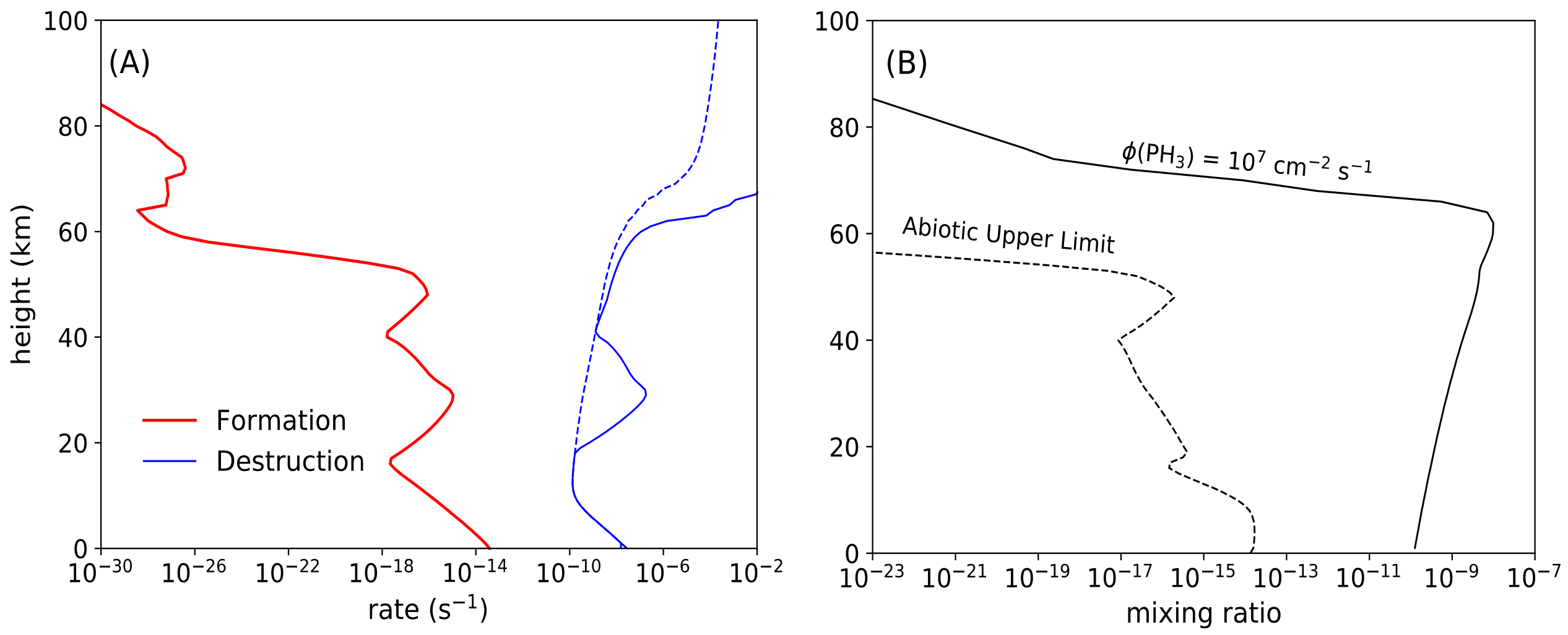

**Fig. 5.** Predicted maximum photochemical production of $PH_3$ (see kinetic network of Supplementary Figure 7), found to be insufficient to explain observations by more than four orders of magnitude. Left panel, (A): Upper limits of the predicted photochemical production rates (excluding transport) (red curve, $s^{-1}$) compared to photochemical destruction rates (blue curve, $s^{-1}$), including radicals and atoms (blue solid) and ignoring radicals and atoms (blue dashed), as a function of height (km). Right panel, (B): Mixing ratio of $PH_3$ as a function of atmospheric height (km), for a production flux within the cloud layer (~55-65 km) of $10^7$ $cm^{-2}$ $s^{-1}$ (solid curve), compared to the predicted steady state abiotic upper limit (dashed curve).

Energetic events are also not an effective route to making phosphine. Lightning may occur on Venus, but at sub-Earth activity levels[33]. We find that $PH_3$-production by Venusian lightning would fall short of few-ppb abundance by factors of $10^7$ or more. Similarly, there would need to be > 200 times as much volcanic activity on Venus as on Earth to inject enough phosphine into the atmosphere (up to ~$10^8$, depending on assumptions about mantle rock chemistry). Orbiter topographical studies suggest there are not many large, active, volcanic hotspots on Venus[34]. Meteoritic delivery adds at most a few tonnes of phosphorus per year (for Earth-like accretion of meteorites). Exotic processes like large-scale tribochemical (frictional) processes and solar wind protons also only generate $PH_3$ in negligible quantities (see Supplementary Figure 10; ref. 35).

## Discussion

If no known chemical process can explain $PH_3$ within the upper-atmosphere of Venus, then it must be produced by a process not previously considered plausible for Venusian conditions. This could be unknown photochemistry or geochemistry, or possibly life. Information is lacking – as an example, the photochemistry of Venusian cloud droplets is almost completely unknown. Hence a possible droplet-phase photochemical source for $PH_3$ must be considered (even though phosphine is oxidised by sulphuric acid). Questions of why hypothetical organisms on Venus might make phosphine are also highly speculative (see "$PH_3$ and hypotheses on Venusian life" in Methods and Supplementary Information).

Quantitatively, we can note that the production rates of ~$10^6$-$10^7$ molecules $cm^{-2}$ $s^{-1}$ inferred above are lower than the production by some terrestrial ecologies, which make the gas[10] at $10^7$-$10^8$ $PH_3$ $cm^{-2}$ $s^{-1}$. Considering also distribution, the phosphine on Venus is at or near temperate altitudes, and is also lacking around the polar caps. It is suggested[36] that the mid-latitude Hadley circulation cells offer the most stable environment for life, with circulation times of 70-90 days being adequate for reproduction of (Earth-analog) microbes. Phosphine is not detected by ALMA above an ~$60^o$ latitude-bound, agreeing within ~$10^o$ with the proposed upper Hadley-cell boundary[37] where gas circulates to lower altitudes. However, further work on diffusion processes (Supplementary Figure 9) is desirable .

In the context of Solar-System biosignature searches, our observations of the $PH_3$ 1-0 line have proved powerful for modest facility time (<10 hours on-source). The phosphine abundance is well-enough constrained (within factors ~2-3) for worthwhile modelling, and no *ad-hoc* introduction of temporal effects is needed. We have ruled out contaminants, and narrow lines mean that a presently-unknown chemical species would need to have a transition at an extremely nearby wavelength to mimic the $PH_3$ 1-0 line. However, confirmation is always important for a single-transition detection. Other $PH_3$ transitions should be sought, although observing higher-frequency spectral features may require a future large air- or space-borne telescope.

Even if confirmed, we emphasize that the detection of phosphine is not robust evidence for life, only for anomalous and unexplained chemistry. There are substantial conceptual problems for the idea of life in Venus' clouds – the environment is extremely dehydrating as well as hyper-acidic. However, we have ruled out many chemical routes to phosphine, with the most-likely ones falling short by 4-8 orders of magnitude (Supplementary Figure 10). To further discriminate between unknown photochemical and/or geological processes as the source of Venusian phosphine, or to determine if there is life in the clouds of Venus, substantial modelling and experimentation will be important. Ultimately, a solution could come from revisiting Venus for *in situ* measurements or aerosol return.

**References:**

1 Baudino, J.-L. *et al.* Toward the analysis of JWST exoplanet spectra: Identifying troublesome model parameters. *The Astrophysical Journal* **850**, 150 (2017).

2 Boston, P. J., Ivanov, M. V. & McKay, C. P. On the possibility of chemosynthetic ecosystems in subsurface habitats on Mars. *Icarus* **95**, 300-308 (1992).

3 McKay, C. P., Porco, C. C., Altheide, T., Davis, W. L. & Kral, T. A. The possible origin and persistence of life on Enceladus and detection of biomarkers in the plume. *Astrobiology* **8**, 909-919 (2008).

4 Pappalardo, R. T. *et al.* Does Europa have a subsurface ocean? Evaluation of the geological evidence. *Journal of Geophysical Research: Planets* **104**, 24015-24055 (1999).

5 Roth, L. *et al.* Transient water vapor at Europa's south pole. *Science* **343**, 171-174 (2014).

6 Waite, J. H. *et al.* Cassini ion and neutral mass spectrometer: Enceladus plume composition and structure. *science* **311**, 1419-1422 (2006).

7 Postberg, F. *et al.* Macromolecular organic compounds from the depths of Enceladus. *Nature* **558**, 564-568 (2018).

8 Oehler, D. Z. & Etiope, G. Methane seepage on Mars: where to look and why. *Astrobiology* **17**, 1233-1264 (2017).

9 Gillen, E., Rimmer, P. B. & Catling, D. C. Statistical analysis of Curiosity data shows no evidence for a strong seasonal cycle of Martian methane. *Icarus* **336**, 113407 (2020).

10 Sousa-Silva, C. *et al.* Phosphine as a Biosignature Gas in Exoplanet Atmospheres. *Astrobiology* **20**, doi:10.1089/ast.2018.1954 (2020).

11 Pasek, M. A., Sampson, J. M. & Atlas, Z. Redox chemistry in the phosphorus biogeochemical cycle. *Proceedings of the National Academy of Sciences* **111**, 15468-15473, doi:10.1073/pnas.1408134111 (2014).

12 Bregman, J. D., Lester, D. F. & Rank, D. M. Observation of the nu-squared band of PH3 in the atmosphere of Saturn. *Astrophysical Journal* **202**, L55-L56, doi:10.1086/181979 (1975).

13 Tarrago, G. *et al.* Phosphine spectrum at 4–5 μm: Analysis and line-by-line simulation of 2ν2, ν2 + ν4, 2ν4, ν1, and ν3 bands. *Journal of Molecular Spectroscopy* **154**, 30-42, doi:https://doi.org/10.1016/0022-2852(92)90026-K (1992).

14 Noll, K. S. & Marley, M. S. in *Planets Beyond the Solar System and the Next Generation of Space Missions.* (ed David Soderblom) 155 (1997).

15 Visscher, C., Lodders, K. & Fegley, B. Atmospheric Chemistry in Giant Planets, Brown Dwarfs, and Low-Mass Dwarf Stars. II. Sulfur and Phosphorus. *The Astrophysical Journal* **648**, 1181 (2006).
16 Morowitz, H. & Sagan, C. Life in the Clouds of Venus? *Nature* **215**, 1259 (1967).
17 Limaye, S. S. *et al.* Venus' Spectral Signatures and the Potential for Life in the Clouds. *Astrobiology* **18**, 1181-1198 (2018).
18 Bains, W., Petkowski, J. J., Sousa-Silva, C. & Seager, S. New environmental model for thermodynamic ecology of biological phosphine production. *Science of The Total Environment* **658**, 521-536 (2019).
19 Weisstein, E. W. & Serabyn, E. Detection of the 267 GHz J= 1-0 rotational transition of PH3 in Saturn with a new Fourier transform spectrometer. *Icarus* **109**, 367-381 (1994).
20 Cram, T. A Directable Modular Approach to Data Processing. *Astronomy and Astrophysics Supplement Series* **15**, 339 (1974).
21 Warmels, R. *et al. ALMA Cycle 6 Technical Handbook*. Vol. ALMA Doc. 6.3 (2018).
22 Encrenaz, T., Moreno, R., Moullet, A., Lellouch, E. & Fouchet, T. Submillimeter mapping of mesospheric minor species on Venus with ALMA. *Planetary and Space Science* **113**, 275-291 (2015).
23 Jacquinet-Husson, N. *et al.*, The 2015 edition of the GEISA spectroscopic database. *Journal of Molecular Spectroscopy* **327**, 31-72 (2016).
24 Gordon, I. E. *et al.* The HITRAN2016 molecular spectroscopic database. *Journal of Quantitative Spectroscopy and Radiative Transfer* **203**, 3-69 (2017).
25 Tennyson, J. *et al.* The ExoMol database: molecular line lists for exoplanet and other hot atmospheres. *Journal of Molecular Spectroscopy* **327**, 73-94 (2016).
26 Kuczkowski, R. L., Suenram, R. D. & Lovas, F. J. Microwave spectrum, structure, and dipole moment of sulfuric acid. *Journal of the American Chemical Society* **103**, 2561-2566 (1981).
27 Piccialli, A. *et al.* Mapping the thermal structure and minor species of Venus mesosphere with ALMA submillimeter observations. *Astronomy & Astrophysics* **606**, A53 (2017).
28 Gurwell, M. A., Melnick, G. J., Tolls, V., Bergin, E. A. & Patten, B. M. SWAS observations of water vapor in the Venus mesosphere. *Icarus* **188**, 288-304 (2007).
29 Dartnell, L. R. *et al.* Constraints on a potential aerial biosphere on Venus: I. Cosmic rays. *Icarus* **257**, 396-405 (2015).
30 Sousa-Silva, C., Hesketh, N., Yurchenko, S. N., Hill, C. & Tennyson, J. High temperature partition functions and thermodynamic data for ammonia and phosphine. *Journal of Quantitative Spectroscopy and Radiative Transfer* **142**, 66-74 (2014).
31 Sousa-Silva, C., Tennyson, J. & Yurchenko, S. N. Communication: Tunnelling splitting in the phosphine molecule. *The Journal of Chemical Physics* **145**, doi: 10.1063/1.4962259 (2016).
32 Krasnopolsky, V. A. Vega mission results and chemical composition of Venusian clouds. *Icarus* **80**, 202-210, doi:https://doi.org/10.1016/0019-1035(89)90168-1 (1989).
33 Lorenz, R. D. Lightning detection on Venus: a critical review. *Progress in Earth and Planetary Science* **5**, 34 (2018).
34 Shalygin, E. V. *et al.* Active volcanism on Venus in the Ganiki Chasma rift zone. *Geophysical Research Letters* **42**, 4762-4769 (2015).
35 Bains, W. *et al.* Phosphine on Venus Cannot be Explained by Conventional Processes. *Astrobiology* (under review, 2020).

36 Grinspoon, D. H. & Bullock, M. A. Astrobiology and Venus exploration. *Geophysical Monograph - American Geophysical Union* **176**, 191 (2007).
37 Sánchez-Lavega, A., Lebonnois, S., Imamura, T., Read, P. & Luz, D. The atmospheric dynamics of Venus. *Space Science Reviews* **212**, 1541-1616 (2017).

Correspondence and requests for materials should be addressed to greavesj1 at cardiff.ac.uk.

**Acknowledgements:** Venus was observed under JCMT Service Program S16BP007 and ALMA Director's Discretionary Time program 2018.A.0023.S. As JCMT users, we express our deep gratitude to the people of Hawaii for the use of a location on Mauna Kea, a sacred site. We thank Mark Gurwell, Iouli Gordon and Mary Knapp for useful discussions; personnel of the UK Starlink Project for software training; Sean Dougherty for award of ALMA Director's discretionary time; and Dirk Petry and other Astronomers on Duty and project preparation scientists at ALMA for ensuring timely observations. The James Clerk Maxwell Telescope is operated by the East Asian Observatory on behalf of The National Astronomical Observatory of Japan; Academia Sinica Institute of Astronomy and Astrophysics; the Korea Astronomy and Space Science Institute; Center for Astronomical Mega-Science (as well as the National Key R&D Program of China with No. 2017YFA0402700). Additional funding support is provided by the Science and Technology Facilities Council of the United Kingdom and participating universities in the United Kingdom (including Cardiff, Imperial College and the Open University) and Canada. The Starlink software is currently supported by the East Asian Observatory. ALMA is a partnership of ESO (representing its member states), NSF (USA) and NINS (Japan), together with NRC (Canada), MOST and ASIAA (Taiwan), and KASI (Republic of Korea), in cooperation with the Republic of Chile. The Joint ALMA Observatory is operated by ESO, AUI/NRAO and NAOJ. Funding for the authors was provided by STFC (grant ST/N000838/1, DLC); Radionet/MARCUs through ESO (JSG); the Japan Society for the Promotion of Science KAKENHI (Grant No. 16H02231, HS); the Heising-Simons Foundation, the Change Happens Foundation, the Simons Foundation (495062, SR) and Simons Foundation SCOL award (59963, PBR). RadioNet has received funding from the European Union's Horizon 2020 research and innovation programme under grant agreement No 730562.

**Author contributions:** JSG and AMSR analysed telescope data; HS developed a radiative transfer model; JJP and WB worked out chemical kinetics and thermodynamics calculations; SR, PR, JJP, WB, SS worked on photochemistry; CS provided spectroscopic expertise and line parameter analysis; AC, DC, EDM, HF, CS, SS, IMW, ZZ contributed expertise in astrochemistry, astrobiology, planetary science and coding; PF, IC, EL and JH designed, made and processed observations at the JCMT. JG, WB, JJP, DC, SS and PR wrote the paper.

**Data availability:** The data that support the plots within this paper and other findings of this study are available from the corresponding author upon reasonable request. The raw data are publicly available at websites https://www.eaobservatory.org/jcmt/science/archive/ (JCMT) and http://almascience.eso.org/aq/ (ALMA).

**Code availability:** Our reduction scripts that can be used to reproduce the astronomical results shown are provided as Supplementary Software 1 (JCMT) and 2-4 (ALMA).

# Methods

## *JCMT data reduction*

The reduction was all performed offline, within the Starlink Project software suite[38]. The manipulation of timestream data used KAPPA[39], the cube was made in SMURF[40], and the final spectra were constructed in SPLAT[41]. The software is open access and is supported by the East Asian Observatory. Our data-processing can be reproduced using the script in Supplementary Software 1.

Standard methods in millimeter-waveband single-dish spectroscopy were used to process 140 JCMT spectra, finally reaching a dynamic range better than $10^4$ in l:c. Ref. 42 describe specifics of observing Venus with the JCMT; we omitted their telluric-correction step as any trace terrestrial phosphine would be supressed by beam-switching and calibration, and offset in velocity by 13.5-14.1 km/s at the time of our Venus observations.

Undulations in amplitude versus frequency, equivalently Doppler-shifted velocity, occur across the passband. Time-delayed signals due to reflections of the bright planetary-continuum (e.g. reflection off the calibration cold-load not seen when observing the sky) yield difference-spectra with residual ripples. Wave-period analysis[43] suggests two major effects from reflections within the receiver cabin (for details, see Supplementary Information: JCMT data reduction). Both effects were removed by fitting polynomial functions, with 'troughs' much broader than the core of the absorption line we seek (Supplementary Figure 1). However, this subtraction removes the line wings, expected to span tens of MHz; after this step, the line-center l:c value under-estimates the true line depth (Figure 1b).

A third ripple has troughs of comparable width to the line and similar amplitude (for origins, see Supplementary Information), sometimes shifting phase within each ~30-minute observation (Supplementary Figure 1). For ~16 waves across a 250 MHz passband (281 km/s at 266.9445 GHz), each half-wave occupies ~9 km/s, so an artefact 'trough' is similar in width to model-lines. We applied polynomial-fits on a sub-scan basis, restricting the passband to avoid use of high orders. Over 100 km/s, 10-11 peaks/troughs are expected, with 1-2 being interpolated across (overlaying the line). The polynomial interpolates over a user-selected region around the line velocity, to avoid fitting (and removing) the line as if it were an artefact. This interpolation-region can not exceed ~18 km/s or the polynomial has no information to detect e.g. a downwards trend between two peaks. The polynomials finally adopted were of $8^{th}$-order, an empirical minimum that reasonably flattened the baseline. This order is just below the expected value of (N+1) = 9-11, for N = 8-10 peaks/troughs outside the interpolation-region.

The possible interpolation-regions within model and data constraints span ~4-16 km/s. Figure 1b adopts $|v|$ = 5 km/s, i.e. fitted at velocities $v < -5$ and $v > +5$ km/s with respect to Venus, and interpolated over -5 to +5 km/s. The alternate $|v|$ provide minimum and maximum solutions to line depth (Fig. 1a). The smallest $|v|$-value of 2 km/s is very strict and only allows the highest-confidence spectral-bins to be detected; this range falls below the half-width half-minimum in our line-model (red dashed curve, Figure 1b). The largest $|v|$-value of 8 km/s is an approximate upper bound, as for larger $|v|$, polynomial-amplitudes become greater in the passband-center than at all other velocities. This 'bulge' appears unrealistic, and can artificially force the line to be deeper and wider. These two approaches bracket line-depths that can be obtained from the data.

For each sub-observation, we calculated [(data+continuum)/(polynomial+continuum)-1], in a standard approach where division by the baseline-fit performs better than subtraction. "Data" is the residual after the first two polynomial-subtractions, "polynomial" is the spectral-baseline in the third fitting-step, and "continuum" is the mean signal before any subtractions. Our script yields 140 sub-observation l:c-spectra plus a co-add weighted by per-observation system temperature. Standard deviations are generated on a per-channel basis from the dispersion of 140 input-values, and agree within ~1% from channel to channel.

### *ALMA data reduction*

Details of data acquisition are given in Supplementary Information. Important steps included choosing local oscillator settings to minimise instrumental effects, and accurately tracking positions of Venus and the calibrator Callisto. Their flux densities (ignoring spectral lines) were set using the NASA ephemeris and the 'Butler-JPL-Horizons 2012' standard[44]. Frequency channels were mapped to velocity channels in processing using the Venus ephemeris, so that in the final image cubes, 0 km/s corresponds to line-frequency in the planetary-motion frame of Venus.

The UK ALMA Regional Centre (ARC) Node retrieved the raw data from the ALMA Science Archive and ran the standard manually-generated calibration script ('QA2' in ALMA quality assurance documentation and Technical Handbook) on each execution block (EB). Calibration and imaging followed normal imaging procedures but with a few modifications:

- For flux scale and bandpass calibration involving Callisto, baselines shorter than 180 klambda (~200 m), just inside the first null, were used, providing at least one baseline for every antenna to the reference antenna.
- Viewing the calibrated visibility spectrum and imaging Venus revealed a spectral ripple, worst on short baselines.The absorption feature is still seen, but these baseline-dependent errors impede accurate image analysis. There was insufficient signal-to-noise on Callisto for baseline-dependent calibration, so we flagged all baselines <33 m at an early stage (choice described in the Supplementary Information).

Absorption is not present if Callisto is reduced in a similar fashion (excluding short baselines; same channel selection for continuum-subtraction from visibilities). The bandpass-correction using Callisto used an 8-MHz channel averaging, i.e. such a narrow line would have remained in Callisto, but this was not seen. Hence, the $PH_3$ signal is associated with Venus, not the calibrator.

The initial calibration scripts used to produce the data-cubes that we analysed are provided as Supplementary Software 2 and 3. After applying calibration, we combined all Venus data for each spectral configuration, here describing processing of the wideband and narrowband configurations centred on 266.9 GHz.

We performed time-dependent phase self-calibration using continuum-only channels and applied the solutions to all channels. We excluded the central channels (and, for wideband, channels covering a telluric line), and subtracted a 1$^{st}$-order fit to the continuum from each visibility dataset.

Separately, wideband data with all baselines present were processed similarly up to applying phase self-calibration, and used to produce a continuum image, diameter 15.36 arcsec (verifying the positional accuracy of the ephemeris and tracking). We used a mask of this size, centred on Venus (Supplementary Figure 2) in imaging the continuum-subtracted narrow-band cube.

ALMA spectra were extracted using CASA. Inspection of a grid of spectra (Supplementary Figure 3) showed a residual ripple, varying in shape at adjacent grid points (regardless of grid spacing). Ripples are especially pronounced for spectra at the planetary limb. Gradients across the field are apparent when viewing individual channels. These artefacts appear to arise from the high brightness of Venus (almost filling the primary beam), as a net spectrum surrounding (but excluding) the planet showed a ripple that is the inversion (opposite amplitude-sign) of the ripple in the whole-planet spectrum.

The best mitigation was to extract net-spectra for bands of latitude across the planet. All such bands were drawn in the CASA viewer (see Supplementary Figure 2), with connection-points estimated as positioned within two 0.16-arcsec pixels, below the restoring-beam scale. This 'side-to-side' (zonal mean) methodology takes advantage of gradients such that ripples partially cancel, and $PH_3$ lines becomes visible over extended regions. Resulting spectra (Figure 2) have least noise around the equator, and are noisiest/most-ripply towards the poles (areas dominated by beams at the limb). Attempts to construct similar bands spanning longitude-ranges did not make spectra that were satisfactorily stable, i.e. they changed significantly with sub-beam shifts of boundary-points. Longitudes at the limb exhibit more spectral ripple, and also gradients along this axis appear not to self-cancel ripples.

The final region-averaged spectra were analysed and plotted in SPLAT[41].

In producing final ALMA spectra (Figure 2), we subtracted polynomial functions as for JCMT data, but only needed to do this once for each region-averaged spectrum. The flattest spectral-baselines were obtained by restricting the passband to ±40 km/s from Venus' velocity, and interpolating across |v| = 5 km/s. Polynomials of 12$^{th}$-order were necessary, higher than the 8$^{th}$-order JCMT fits; we verified that lower orders, e.g. 11, yielded worse noise while higher orders, e.g. 14, did not improve the noise. The spectra are shown in Supplementary Figure 4 with the fits overlaid, to demonstrate that the $PH_3$ lines are distinct and unlikely to be subtraction-artefacts.

As a robustness check, a preliminary reduction of the HDO transition observed simultaneously is shown in Supplementary Figure 5. (Prior to a full investigation, we have only spectrally-binned the narrowband data to flatten ripples and applied a 1$^{st}$-order polynomial-fit.) The same processing was applied to a model line from our radiative transfer code, for an HDO-abundance set to 2.5 ppb (red curve). If HDO is smoothly distributed, this line will suffer losses from the missing ALMA short baselines, and so for losses e.g. comparable to the continuum signal, up to ~14 ppb of HDO is possible. The $H_2O$ abundance would then be ~0.04-0.2 ppm, following ref. 22 for deuteration-correction. This is within low-end water abundances reported from millimetre-waveband monitoring – values are strongly time-variable, with detections at ~0.1-3.5 ppm[45].

We verified $PH_3$ absorption in wideband data, which are simultaneous (coming from the same photons). These spectra have substantial ripple (Supplementary Figure 6), so are only minimally-processed, via a 1$^{st}$-order polynomial-fit using velocity-ranges of -20 to -5, +5 to +20 km/s. In the wideband data, the phosphine line is present but only of modest significance. The line minimum is ~50% deeper than in the narrowband data, with velocity-centroid at ~-0.9±0.6 km/s.

***Calculation of $SO_2$ contamination***

We also used the ALMA wideband-tunings to search for other transitions, especially checking whether our phosphine line-candidate could be a mis-identification of a sulphur dioxide feature. This $SO_2$ transition (Supplementary Table 1) would comprise absorption from a lower-level J=31 rotational state. This energy-state is at 613 K, and our wideband spectra also covered transitions from J=13, 28 and 46, at energy-levels of 93, 403 and 1090 K respectively. In principle, an excitation diagram of these three lines could yield an interplated flux of the J=31 “interloper”, but in practice none of the transitions was detected. $SO_2$ lines from Venus even at energy levels ~100 K are challenging, e.g. noting 1/4 successful attempts in ref. 22 (with only half the antennas commissioned for their early-ALMA run).

In our whole-planet wideband spectrum, we do not detect the most favorable $SO_2$ transition. We adopt a very generous upper-limit on this line, allowing ~100% of peak-to-peak amplitude of nearby ripples (Figure 4). In our radiative transfer model, this limit requires $SO_2$ abundance of 10 ppb at ≥88 km, in reasonable agreement with ~16 ppb fitted to a prior mesospheric-altitudes detection with ALMA[27] (and without attempting line-dilution corrections). Our model then predicts l:c < $2{\cdot}10^{-5}$ for the ‘interloper’ J=31 line-minimum. After polynomial-fitting to this model as for the observations, l:c integrated over ±5 km/s is only 8% of the feature identified as phosphine. After subtracting the maximised $SO_2$ contamination (Figure 4), the velocity centroid of the residual moves towards zero as might be expected if the phosphine identification is correct, but by <0.1 km/s. Figure 4 also demonstrates that $SO_2$ J=31 cannot reproduce the *entire* line identified as $PH_3$, as the $SO_2$ J=13 component would strongly violate the wideband limit.

### *Abundance retrieval*

Radiative transfer in the Venus atmosphere is calculated assuming spherically-homogeneous atmospheric layers. The atmosphere from the surface up to 140 km is divided into 1-km depth layers, with a temperature profile from a Venus international reference model (VIRA), established from spacecraft temperature measurements. A vertically-constant $PH_3$ mixing ratio is assumed here. Atmospheric opacity at each Venus altitude (Supplementary Figure 2) is calculated with line-by-line computation of molecular lines (CO, $SO_2$, $H_2O$, $PH_3$…) and continuum-opacity due to the collision-induced absorption of $CO_2$. Scattering due to cloud particles is omitted as negligible in our millimeter-waveband. The ‘emission angle effect’ (longer line-of-sight at the limb of Venus) is included. Pencil-beam-spectra are calculated with a fine-grid along the radial direction of the Venus disk, accounting for emission angle, and then convolved with ALMA’s synthesized beam (an elliptic Gaussian). Finally, spectra within the apparent disk are all integrated to obtain a disk-averaged model spectrum. For the JCMT simulation, the process was the same but it used the JCMT gaussian beam size instead of ALMA’s synthesis beam.

The largest source of uncertainty in the retrieved abundances is a lack of collisional broadening parameters for the $PH_3$ J= 1-0 transition. Ammonia is often used as proxy to phosphine as the two molecules share many similarities, including comparably dense energy levels. In the absence of measurements for $CO_2$ broadening/shift effects on the $PH_3$ J 1-0 line, we considered two alternatives: (a) measured $CO_2$ broadening effect of the $NH_3$ J 1-0 line (0.2862 $cm^{-1}$/atm), and (b) a theoretical value estimated from a ratio of known broadening of $H_2$/He for both $PH_3$ and $NH_3$ (0.186 $cm^{-1}$/atm).

### *Photochemical model*

Within the atmosphere of Venus, $PH_3$ is destroyed by photochemically-generated radical species, by near-surface thermal decomposition, and by photodissociation within/above the clouds. Since $PH_3$ itself scavenges chemically-reactive radicals and atoms, e.g. OH, H, O, and Cl, the presence of $PH_3$ suppresses these species, increasing its lifetime. Previously published models of Venus' atmosphere did not include the scavenging effect of $PH_3$, so we developed our own model.

We employ the 1D photochemistry-diffusion code ARGO[46] to solve the atmospheric transport equation. For the UV-transport, a 'mysterious absorber' is included[47]. For our chemical network, we use STAND2019[48], which includes H/C/N/O species, adding a limited S/Cl/P network relevant for Venus by adapting published low- and middle- atmosphere networks[47,49,50] and developing an estimate for thermal decomposition. The Supplementary Information discusses additional reactions and constants[51-55] and major uncertainties. The network includes 460 species and 3406 forward reactions; see ref. 35.

We follow refs. 47,57 in temperature-pressure (TP) profiles from VIRA. TP profiles are from ref. 58, for the deep atmosphere profile (0-32 km) and for altitudes 32-100 km (45 degrees latitude profile adopted), while for altitudes 100-112 km we use the VIRA dayside profile[59]. We followed refs. 47,57 in the Eddy diffusion profile, taken to be constant at $2.2 \cdot 10^3$ $cm^2$ $s^{-1}$ for z <30 km, $1 \cdot 10^4$ $cm^2$ $s^{-1}$ for 47-60 km, $1 \cdot 10^7$ $cm^2$ $s^{-1}$ for >100 km, and connected exponentially at intermediate altitudes. Supplementary Figure 8 shows the temperature-pressure and Eddy diffusion profiles adopted in this work, along with decomposition timescales.

We take fixed[47] surface boundary conditions for the major atmospheric species, along with initial surface boundary conditions for minor species, radicals and atoms, except for $PH_3$. Observational constraints and initial surface model-abundances are listed in Supplementary Tables 2 and 3. In validating our model, predictions for a variety of species compared to observations are plotted in Supplementary Figure 9. Nearly all profiles agree with observations within an order of magnitude in concentration, within 5 km height; for full details, see Supplementary Information.

Supplementary Figure 9 also summarises wind data in relation to our observations of Venus. The latitudinal $PH_3$-variation is somewhat unexpected, implying that (a) the $PH_3$ observed is largely above 60 km *and* that this work *and* the models presented in this work and in refs 47,49 correctly predict or under-predict the above-cloud concentrations of OH, H, O and Cl, or (b) that there is some unknown mechanism that more rapidly destroys $PH_3$ within the cloud layer, or both.

***Potential pathways for phosphine production.***

Potential $PH_3$-production pathways in the Venusian environment are discussed in detail in the Supplementary Information (summarised in Supplementary Figure 10; see also ref. 35). Two possible classes of routes for the production of phosphine were considered: photochemical production or non-photochemical chemistry.

For photochemical modelling, we created a network of reactions of known kinetic parameters[60] that could lead from $H_3PO_4$ (phosphoric acid) to $PH_3$ (phosphine), by reaction with photochemically-generated radicals in the Venusian atmosphere. Where reactions were possible but no kinetic data for the phosphorus species was known, homologous nitrogen species reaction kinetics were used instead, validated by comparing reactions of analogous nitrogen and phosphorus species. The maximum possible rate for reductive chemistry in this network were compared to the destruction rate as a function of altitude.

Non-photochemical reactions were modelled thermodynamically. For surface and atmospheric chemistry, we created a list of chemicals, their concentrations, and reactions, for all potential $PH_3$ production reactions. Phosphorus species abundances were calculated themodynamically and assumed to be in equilibrium with liquid/solid species at the cloud base. The free energy of reaction, indicating whether the net production of phosphine was thermodynamically favoured, was calculated using standard methods (See Supplementary Information for details). None of the reactions favour the formation of phosphine, on average having a free energy of reaction of +100kJ/mol (Supplementary Figure 7).

Modelling the subsurface chemistry was approached via oxygen fugacity ($fO_2$)[61], the notional concentration of free oxygen in the crustal rocks. We model the equilibrium between phosphate and phosphine, for temperatures between 700K and 1800K, at 100 or 1000 bar and with 0.01%, 0.2% and 5% water. Oxygen fugacity of plausible crust and mantle rocks based on Venus lander geological data is 8-15 orders of magnitude too high to support reduction of phosphate, so degassing of mantle rocks would produce only trivial amounts of phosphine. Volcanic, lightning and meteoritic delivery were calculated based on parallels with terrestrial rates of events within Venusian atmosphere, and were calculated to be negligible.

### ***$PH_3$ and hypotheses on Venusian life***

In the SI, we briefly summarise ideas on why the temperate but hyper-acidic Venusian clouds have been proposed for decades as potentially habitable, in spite of obvious difficulties such as resisting destruction by sulphuric acid. We earlier proposed that any detectable phosphine found in the atmosphere of a rocky planet is a promising sign of life[10] and showed that biological production of phosphine is favored by cool, acid conditions[18]. Initial modelling based on terrestrial biochemistry suggests that biochemical reduction of phosphate to phosphine is thermodynamically feasible under Venus cloud conditions[35]. We have also described a possible life-cycle for a Venusian aerial biosphere[62].

**Methods References:**

38 Currie, M. J. *et al.* Starlink Software in 2013. *Astron. Soc. Pacific Conf. Ser.* **485**, 391-394 (2014).

39 Currie, M. J., Berry, D. S. KAPPA: Kernel Applications Package. Astrophysics Source Code Library, ascl:1403.022 (2014).

40 Jenness. T. *et al.* Automated reduction of sub-millimetre single-dish heterodyne data from the James Clerk Maxwell Telescope using ORAC-DR. *Mon. Not. R. Astron. Soc.* **453**, 73-88 (2015).

41 Škoda, P., Draper, P. W., Neves, M. C., Andrešič, D., Jenness, T. Spectroscopic analysis in the virtual observatory environment with SPLAT-VO. *Astron. Comput.* **7**, 108-120 (2014).

42 Sandor, B. J. & Clancy, R. T. First measurements of ClO in the Venus atmosphere– altitude dependence and temporal variation. *Icarus* **313**, 15-24 (2018).

43 Barnes, D. G., Briggs, F. H. & Calabretta, M. R. Postcorrelation ripple removal and radio frequency interference rejection for Parkes Telescope survey data. *Radio Sci.* **40**, 1-10 (2005).

44 Butler, B. ALMA Memo 594: Flux Density Models for Solar System Bodies in CASA. *ALMA Memo Series. NRAO, Charlottesville, VA* (2012).

45 Sandor, B. J. & Clancy, R. T. Water vapor variations in the Venus mesosphere from microwave spectra. *Icarus* **177**, 129-143 (2005).
46 Rimmer, P. B. & Helling, C. A chemical kinetics network for lightning and life in planetary atmospheres. *Astrophys. J. Suppl. Ser.* **224**, 9 (2016).
47 Krasnopolsky, V. A. Chemical kinetic model for the lower atmosphere of Venus. *Icarus* **191**, 25-37 (2007).
48 Rimmer, P. B. & Rugheimer, S. Hydrogen cyanide in nitrogen-rich atmospheres of rocky exoplanets. *Icarus* **329**, 124-131 (2019).
49 Krasnopolsky, V. A. S3 and S4 abundances and improved chemical kinetic model for the lower atmosphere of Venus. *Icarus* **225**, 570-580 (2013).
50 Zhang, X., Liang, M. C., Mills, F. P., Belyaev, D. A. & Yung, Y. L. Sulfur chemistry in the middle atmosphere of Venus. *Icarus* **217**, 714-739 (2012).
51 Burcat, A. & Ruscic, B. Third millenium ideal gas and condensed phase thermochemical database for combustion (with update from active thermochemical tables). (Argonne National Lab.(ANL), Argonne, IL (United States), 2005).
52 Visscher, C. & Moses, J. I. Quenching of carbon monoxide and methane in the atmospheres of cool brown dwarfs and hot Jupiters. *Astrophys. J.* **738**, 72 (2011).
53 Lyons, J. R. An estimate of the equilibrium speciation of sulfur vapor over solid sulfur and implications for planetary atmospheres. *J. Sulphur Chem.* **29**, 269-279 (2008).
54 Kulmala, M. & Laaksonen, A. Binary nucleation of water–sulfuric acid system: Comparison of classical theories with different H2SO4 saturation vapor pressures. *J. Chem. Phys.* **93**, 696-701 (1990).
55 Bierson, C.J. and Zhang, X. Chemical cycling in the Venusian atmosphere: A full photochemical model from the surface to 110 km. *J. Geophys. Res. Planets* doi:10.1029/2019JE006159 (2019).
56 Sander, R. Compilation of Henry's law constants (version 4.0) for water as solvent. *Atmos. Chem.Phys.* **15** (2015).
57 Krasnopolsky, V. A. A photochemical model for the Venus atmosphere at 47–112 km. *Icarus* **218**, 230-246 (2012).
58 Seiff, A. *et al.* Models of the structure of the atmosphere of Venus from the surface to 100 kilometers altitude. *Adv. Space Res.* **5**, 3-58 (1985).
59 Keating, G. M. *et al.* Models of Venus neutral upper atmosphere: Structure and composition. *Adv. Space Res.* **5**, 117-171 (1985).
60 Linstrom, P. J. & Mallard, W. G. The NIST Chemistry WebBook: A Chemical Data Resource on the Internet. *J. Chem. Eng. Data* **46**, 1059-1063, doi:10.1021/je000236i (2001).
61 Frost, B. R. in *Reviews in minerology vol 25. Oxide minerals: petrologic and magnetic significance.* (ed Donald H. Lindsley) (Mineralogical Society of America, 1991).
62 S. Seager, J. J. Petkowski, P. Gao, W. Bains, N. C. Bryan, S. Ranjan, J. Greaves. The Venusian Lower Atmosphere Haze as a Depot for Desiccated Microbial Life: a Proposed Life Cycle for Persistence of the Venusian Aerial Biosphere. *Astrobiology* https://doi.org /10.1089/ast.2020.2244 (2020).

# Supplementary Information;

## Phosphine Gas in the Cloud Decks of Venus

### <u>Astronomical observations</u>

#### ***Details of the JCMT data acquisition and reduction approach***

The James Clerk Maxwell Telescope is a 15m dish on Mauna Kea, Hawaii. All JCMT observations are available in the public archive at CADC, under project id S16BP007 (a Service Programme approved in semester 2016B), and were made with the RxA3 receiver and the ACSIS correlator. Full details are listed here to facilitate independent analysis of the raw data.

The earliest observations only established the best sideband tuning and the additional length of time needed in calibration (to avoid correlated noise from the bright planetary signal). The usable observations were taken on mornings of June 2017: the 9$^{th}$ (observation #49), 11$^{th}$ (#96,97), 12$^{th}$ (#97,100), 14$^{th}$ (#76,78) and 16$^{th}$ (#49,52,55,59). Each observation had 14 samples of ~2 min duration dumped by ACSIS (except that #97,100 on 12$^{th}$ June comprise one standard observation split into two parts). Hence the total number of independent spectra obtained is 140. The ACSIS setup provided 8192 spectral channels across a 250 MHz passband, resulting in channel-widths equivalent to 0.03427 km/s.

One other observation (#79 on 10$^{th}$ June) has been rejected from the analysis as it was excessively noisy, possibly due to an accidental short calibration. A number of shorter observations also appear in the archive, that were terminated by the telescope operator due to a receiver lost-lock report. Some of these were possibly false alarms, but we rejected all these short observations, which would add only 10% to the total time spent observing Venus.

At the time of observation, Venus was just over half-illuminated (phase of 0.55) and subtended 22 arcseconds. Hence the limb is downweighted by ~50%, given the approximately 19 arcsec full-width half-maximum of the telescope beam. The observations were made by beam-switching (chopping the telescope's secondary mirror) off Venus by 60 arcsec to remove the terrestrial sky signal. All ten observations over the five observing days are comparable in quality, with mean and standard deviation in system temperature of 460 K and 50 K respectively. The sky opacities at 225 GHz ranged from 0.09 to 0.17 (from simultanous JCMT Water Vapor Radiometer data) and the airmasses towards Venus ranged from 1.01 to 1.90. Data were taken at UT from 15:30 to 20:40 (observation mid-points), corresponding to 05:30 to 10:40 in local (Hawaii) time.

The observations started later than intended (in semester 2017A), as they were delayed by the calibration procedure being established. As a result, Venus' recession from the Earth caused it to underfill the telescope beam, and the continuum signal (expressed in antenna temperature units, $T_A$*) is less than the brightness temperature of Venus at 266.9445 GHz. The mean continuum signal over the observations was $T_A$* = 152 K, with standard deviation of 15 K. The continuum strength also varied significantly between some sub-observations, even though only ~2 min apart, probably due to imperfect telescope tracking. This could reduce sensitivity but has no effect in our line analysis, as a line-to-continuum (l:c) spectrum was created for each sub-observation.

Two dominant sources of spectral wave-effects (ripples) exist. The primary wave may be from reflection between the main dewar and cold-load dewer (distance ~0.5 m) and the secondary wave from an unidentified surface near the middle of the receiver cabin (of diameter a little over 3 m), with these distances derived from a relation in ref. 43. The dominant ripple is quasi-sinusoidal, with amplitude up to ±2 Kelvin in antenna-temperature $T_A$*. This was fitted with a $4^{th}$-order polynomial across the passband (better matching the data than a pure sine function). After subtracting this component, a second much fainter quasi-sinusoid was then removed. Because of its low amplitude (up to ~0.1 K, comparable to possible line strengths), this step was implemented by average-filtering the spectrum over a moving-box of 1024 channels, fitting a $9^{th}$-order polynomial to the filtered spectrum, and then subtracting this polynomial from the unfiltered data.

After subtraction of these components, a third residual ripple remains. By Fourier analysis of the 2-D dataset (time versus spectral channel; Supplementary Figure 1), this third component has periods that cluster around 8 and 16 waves fitting across the passband. These "$2^n$ waves" can result when signal spikes have been processed during correlation. However, at the JCMT there is a known "16 MHz ripple" from reflection between the receiver and the secondary mirror unit, and/or due to cables between the receiver cabin and Nasmyth switch. Since 16 periods of 16 MHz would occupy 256 MHz, and the correlator was configured for a 250 MHz band, these reflections are likely the cause of the 16-period ripple we see (with the 8-period features probably being a modulation).

In principle, the whole 2-D dataset could be collapsed down on the time axis, and then a single polynomial could be fitted to remove the spectral ripple. The data were independently reduced in this way by one of the co-authors, and the phosphine line was recovered. However, we chose the three-step method as optimum for signal-to-noise, relation to physical effects, and ability to generate per-channel noise and check for any peculiar sub-observations (none were identified).

Fourier transform analysis was additionally performed, within KAPPA. There appears to be insufficient signal-to-noise that an FT could capture the phase-drifts shown in Supplementary Figure 1, and so we could not remove the ripples in Fourier space. In particular, the amplitudes of the main FT components were only about half those amplitudes seen in a median filter of the data. The observations have up to six significant features in Fourier space (defined as peaks > $5\sigma$ in the Hermitian-transform image), which results in ripple features not repeating within the passband. This also prevented us from using a fitting method based on a ripple 'template', i.e. a known baseline structure established away from Venus' velocity, that could be applied to the region covering the phosphine line.

The removal of the line wings in the reduction means that the line-center l:c value somewhat under-estimates the true line depth. We processed our model line in the same way as the data (Figure 1b), fitting an $8^{th}$-order polynomial outside the specified line region. This approach is difficult to perform fully realistically, as the ripple artefacts do not repeat within the passband, and so we could not inject such systematics appropriate to the Venus velocity into the model. Some additional tests were made combining our model line with noise plus sine-wave ripples of appropriate periods, and for |v| = 8 km/s the line-minimum is then seen at ~60% of its true value. However, there is significant dispersion, with e.g. a few percent of cases recovering the full line depth, in 1000 tests with different seeds for the random sine waves. In the main text, we adopt the simpler method of processing the noiseless model in our adopted |v| = 5 km/s case, which has also resulted in l:c of ~60% of the original model depth. Our 20 ppb solution after this processing (Figure 1b) fits the observed spectrum well in both line depth and width.

***Details of the ALMA data acquisition and reduction approach***

The Atacama Large Millimetre/submillimetre Array (ALMA) is located on a high plateau site in Chile. A total of forty-four 12m antennas were used, giving baselines of 15-314 m in a compact configuration of ALMA. Our data were obtained under Director's Discretionary Time, as project 2018.A.00023.S. Venus was observed on 2019 March 5 in two contiguous execution blocks (EBs) between 12:19 and 15:50 UTC (between 08:19 and 11:50 in Chilean Standard Time). Compared to the JCMT observation, Venus was somewhat more distant when viewed with ALMA (subtending 15 arcsec) and was also consequently more illuminated (phase of 0.74). Callisto was observed as the bandpass and flux-scale calibrator and J2000-1748 as the phase reference source. System temperatures in the spectral windows (spw) covering the line were around 100 K (between 80-160 K depending on antenna) with no telluric features in the high-resolution spw (the nearest being located at 267.3 GHz). Sky temperatures were 31-34 K in the direction of Venus, and receiver temperatures were 46-51 K. Atmospheric transmission in the high-resolution spw was in the range 89-91%, and precipitable water vapour (PWV) was 1.6 to 2.1 mm, at the lower range of the standard conditions for observing in ALMA Band 6.

We used four frequency bands in two pairs; one of each pair of width 1.875 GHz and the other of 0.1171875 GHz, each divided into 1920 spectral channels. One pair was centred on ~266.9 GHz and the other on ~266.1 GHz. The exact frequency was set at the start of each block to rest frequencies of 266.944662 and 266.16107 GHz, corrected for the motion of Venus. These are the rest frequencies of our $PH_3$ and HDO transitions (but noting that we updated the $PH_3$ rest-frequency in data-analysis, see below). In order to minimise instrumental effects, each frequency band was observed with three different combinations of local oscillator settings (as recommended by ALMA staff), resulting in three spectral windows (spw) for the same final frequency width and resolution. This mitigates the effect of the brightness of Venus on the bandpass dyamic range. We then reduced the effects of stochastic noise by averaging by approximately 8 channels to achieve the velocity resolution of 0.55 km s-1 used in analysis, and reduced the effects of residual ripples by baseline fitting after the extraction of spectra, as described in Methods.

Standard bandpass calibration could not remove ripples due to the large angular size of Venus being poorly sampled by the shortest baselines and to polarization-dependent primary beam effects, apparent for a planet of a few-thousand Jansky (Jy). These ripples were dependent on baseline length, not antenna. After experimenting with flagging baselines <20 to <50 m, and selective flagging of specific baselines, we found that excluding baselines <33 m gave a balance between the amount of ripple produced by shorter baselines and the increase in noise due to removing baselines becoming greater than the residual ripple amplitude. This means that the largest angular scale (LAS) imaged reliably (detected) is 4.3 (7.1) arcsec. According to the current ALMA definition of LAS, the baselines >33m correspond to 5.6 arcsec but imaging will be less reliable close to this scale for narrow spectral channels.

In all calibrations, we inserted the 'Butler-JPL-Horizons 2012' model for Venus. The phase scatter of the calibrated phase reference data is ~5 degrees. Each phase reference scan is ~25 sec. Venus was phase-only self-calibrated (excluding the central channels where the line is expected) using a 30s solution interval. The Venus phase solutions have a similar scatter although the corrected phases are dominated by the structure of Venus. The diameter of Venus produces at least 17 nulls in our uv range, so there was insufficient signal-to-noise for amplitude self-calibration of all antennas, despite hundreds-of-Jy fluxes on the shortest baselines.

Using baselines longer than 33 m, the synthesised beam was 1.088 × 0.777 arcsec, at position angle PA = -88.5°. The image used for analysis was made (see Supplementary Software 4) with a spectral resolution of 0.55 km/s, to reduce the effects of stochastic noise and to match binned JCMT spectra. The ALMA rms near the field-centre is ~3.5 mJy/beam per channel. In all cases, the primary beam correction is applied; the edge of Venus corresponds to about 70% sensitivity.

***ALMA data analysis: additional checks***

In case artificial lines could be produced by applying too high an order in polynomial-fitting, we repeated the data-processing steps of self-calibration and continuum subtraction, but instead of excluding the 400 channels around the band centre, where the line was expected, we performed two alternative reductions excluding 400 channels below and above this central spectral region. We then imaged these channels and analysed them to produce whole-planet spectra, and the outputs are shown in Supplementary Figure 4. The l:c line-integrated signals are 14-22% of the value for the real line, and only ~2 channels in each case appear significant, i.e. the artefacts are narrow. Lorentzian fits to these artefacts give FWHM only half that of the real line (1.8, 2.4 km/s versus 4.1 km/s). This test shows that polynomial fitting does *not* artificially produce features that have profiles like the real $PH_3$ line predicted by realistic radiative transfer models of the atmosphere. To check the robustness of the $PH_3$ line centroid velocities (Table 1), we also used as an alternative 1$^{st}$-order polynomial (linear fits) connecting spectral sections spanning ~5 km/s immediately adjacent to each side of the absorption feature, and then calculated centroids over ±3 km/s ranges. The wideband data provide a consistent but lower-quality estimate of the centroid. These data also in principle include the line wings, but these are obscured by ripple.

All ALMA l:c values have been derived using continuum flux densities of the relevant regions on the planet. These fluxes were measured separately from a continuum data product (imaged with all baselines). This aspect differs from the JCMT products, where each sub-observation contains its own continuum signal. The respective ALMA continuum fluxes are 16.1 (whole planet), 16.3 (equatorial zone), 16.2 (mid-latitudes) and 13.8 (polar caps) Jy/beam. Standard deviations are ~0.1-0.2 Jy/beam, so the fluxes averaged across these areas will more precise than this. No flux differences between northern and southern hemispheres were identified at this precision, but there may be small longitudinal variations (hence possibly with Sun-angle) at ~±0.1 Jy/beam levels.

The continuum image has good fidelity (for example, demonstrating some expected cooling at the poles), and has a whole-planet primary-beam-corrected flux of 2769 Jy at 266.9 GHz. This is 110% of the flux of 2508 Jy on the observing date estimated in the online SMA planetary visibility tool. This difference is moderate given uncertainties in absolute flux calibration, e.g. from the atmospheric temperature assumed in VIRA, and from ~5-7% uncertainty in transferring the flux scale from Callisto. However, the (narrowband) line-fluxes are significantly reduced by the omission of baselines of length <33 m, a reduction step which was not needed for the continuum dataset. This disparity makes it strongly preferable to use the JCMT spectra for comparison to models, as line and continuum signals are obtained from the same JCMT data-product.

We also simulated the ALMA line-dilution problem using a signal of l:c = $-2{\cdot}10^{-4}$ in a circular 'cloud' of 8 arcsec diameter that was placed off-centre in the field, as if towards the planetary limb. Processing this simulation through CASA, we found the line signal was reduced by 37.5% for the whole cloud, if all telescope baselines were used. However, omitting baselines <33 m as in the real processing, 80-90% of the line signal was missing, for different parts of the cloud. Hence the ALMA line-to-continuum ratios suffer from significant line-loss in the case of few-arcsec scale smooth distributions, and under-estimate the true absorption depth. Given that we do not know the actual scales of cloud(s) of phosphine on the real planet, it is difficult to apply any corrections. In the main text, we discuss the maximum correction if the gas distribution is as smooth as the continuum – this structure is highly uniform, being only slightly fainter around the poles (Supplementary Figure 2).

As described in Methods, we verified (at modest significance) that the $PH_3$ absorption was reproduced in the wideband data, which arise from the same photons as the narrowband data. This analysis was made only to verify that the phosphine detection is not an artefact of our spectral-extraction processing. The original intent was to use the wideband data to measure line wings, but this could not be carried out due to the substantial spectral ripple.

## <u>Spectral contamination and uncertainties</u>

### ***Spectral line contaminants additional to $SO_2$***

We checked whether absorption near 266.9445 GHz could be caused by another molecule within the Venus atmosphere. Supplementary Table 1 contains a list of all the molecular candidates within a ±80 MHz range of the $PH_3$ J 1-0 line – thus including transitions that might be detectable in our passbands even if they are not possible contaminants, and also allowing a check for species where transition frequencies have large uncertainties. The most likely contaminant for the $PH_3$ J 1-0 transition is the J=31 $SO_2$ line that we have ruled out as significant.

All other molecules that are known to be present in Venus are not spectrally active within the bandpass, and as such cannot be responsible for any absorption. The molecules in this category that we considered comprise: $CO_2$, $N_2$, $H_2O$, HDO, CO, $H_2SO_4$, $DHSO_4$, OCS, $H_2S$, $O_3$, HCl, DCl, SO, HF, DF, NO. Additional compounds of elements known to be present were also checked – phosgene; disulfur dioxide and isotopologues; other Cl, N, S, species including $ClSO_2$, ClOS, ClCN, HSD, HNSO, HOCl, DOCl, ClNO, NSCl, $ClO_2$, DNO, DCN. All of these had no transitions near our passband or only have data on their infrared transitions.

We also considered other molecules with transitions near the $PH_3$ 1-0 line, but whose presence in the Venusian atmosphere would be equally extraordinary as phosphine. All known transitions that lie within 2 MHz of the $PH_3$ 1-0 rest-frequency are listed in Supplementary Table 1 (upper section) – all the species apart from $PH_3$ and $SO_2$ are highly chemically implausible.

Some of these listed species may produce detectable astronomical signals in other contexts, e.g. envelopes of evolved stars and the interstellar medium. Two examples are an $HC_3N$ line offset by +1.3 km/s with respect to $PH_3$ 1-0, plus an SiS line that is further offset (at +3.1 km/s, just outside the tabulated range). The former line would be discrepant in velocity by $\sim 4\sigma$ and the latter would be displaced into the wings of the phosphine line. $HC_3N$ is a nitrile requiring high energy to make; it would form further bonds to the C-atoms at Venus' atmospheric density; it is probably unstable to UV photolysis; and it has the same problem as $PH_3$ in being a reduced (H-bearing) gas species in an oxidised atmosphere. SiS gas would condense out at any temperature in Venus atmosphere, and also combine with trace water to form solids including $SiO_2$. Both are thus extremely improbable species – posing more problems to form that does $PH_3$. Further, these transitions are both from vibrationally-excited states, from energy levels at approximately 1500, 4300 K ($HC_3N$, SiS respectively) so the levels would be poorly populated in gas at < 300 K.

***Strategies to confirm $PH_3$***

Our recommendation is that phosphine transitions beyond the J=1-0 line be observed in the Venus atmosphere, so that its presence can be confirmed. This is unfortunately very difficult from ground-based facilities, as most transitions lie in challenging wavebands suffering from strong terrestrial atmospheric absorption. The J=3-2 transition of phosphine lies in ALMA Band 10, but it would require ~3 days of observation in first-octile conditions to achieve a 3$\sigma$ statistical significance, for a ~1 km/s spectral bin at line minimum and assuming l:c ~2.5 $10^{-4}$ (possibly an over-estimate, as the continuum will arise higher up at this frequency, potentially reducing the absorbing column above). This experiment would also be challenged by varying terrestrial conditions and observing parameters for Venus over the large number of days needed in practical operations. Other rotational transitions probably require a larger air- or space-borne telescope than are currently available.

Phosphine is also a strong absorber in the infrared, so ground-based telescopes capable of M- and N-band spectroscopy could in principle detect a portion of phosphine's strongest feature at 4-4.8 micrometres, and its broad band centered at 10 micrometres. We searched online archives of large infrared telescopes, specifically VIRTIS/VEX; IRTF (via the IRSA database) for the SpEX and ISHELL instruments plus TEXES (from private communications); and the ESO, Gemini and Subaru archives for all instruments, but did not find any relevant data for Venus. Further, the 4-4.8 micrometre region is likely to be completely dominated by $CO_2$ features. The most sensitive telescopes such as the James Webb Space Telescope (JWST) would have severe saturation and Sun-avoidance issues, so we recommend further investigation of ground-based instruments, possibly with adaptations to observing modes. We caution that mid-infrared and far-infrared observations might *not* detect any phosphine in absorption, depending on the height where the quasi-continuum signal is generated in these wavebands; e.g. at high altitudes, the molecules are expected to be photo-destroyed (Figure 5).

***Adopted line frequency and sources of error in velocities in spectra***

We adopt a line rest frequency for $PH_3$ 1-0 of 266.9445 GHz. For the telescope configurations, we used an older frequency, 266.944662 GHz, but we corrected to the more recently-established rest-frame reference frequency in the data analysis. As a result, the central channel in each spectrum is not exactly at zero velocity of the center of Venus, but is offset by ~0.18 km/s. We neglect any resulting systematic in the calculation of centroid velocities of the lines (i.e. from the outcome that the velocity range has to be slightly asymmetric about zero, for integer channel numbers). Frequency scales of the correlators have a precision better than 0.1 km/s, and are treated as perfect in our analysis. There is no known value for a pressure-related (collisional) shift of the $PH_3$ 1-0 rest frequency – values stated in catalogues are zero, but this is representing unknown rather than negligible. This shift will be orders of magnitude less than pressure broadening effects. Given the signal-to-noise in the spectra, we did not attempt to fit line profiles with asymmetries representing collisional effects at moderate gas-pressure, such as Van Vleck-Weiskopf profiles. Such profiles were however included in the radiative transfer calculations.

ALMA observations were made in the reference-frame of Venus' center with respect to the observatory, but JCMT observations, while tracking the planet's position, adopted zero velocity as being in the rest frame of the telescope. Venus' velocity relative to JCMT was tabulated (from the JPL Horizons tool) for the mid-point of each ~30 minute observation, and each observation was shifted to the Venus frame in analysis (see values in Supplementary Software 1). The line was located at +13.49 to +14.06 km/s in the JCMT telescope-referenced spectra (mean of +13.8 km/s), sufficient to shift the absorption feature away from any terrestrial line at v = 0 km/s, in the highly unlikely event of a small terrestrial plume of phosphine entering the on-source telescope beam. The uncorrected velocity-shifts during each of the JCMT observations are negligible (<0.05 km/s). For the ALMA observations, the relative velocity of Venus was in the range +11.6 to +12.0 km/s, so again no terrestrial contamination is plausible. (There are also no catalogued transitions at a suitable frequency that could mimic phosphine on Venus after applying the velocity-shift between the two planets.) The rotation of Venus' upper atmosphere at ~0.1 km/s cannot be detected within the signal-to-noise of our spatially-resolved ALMA data.

## Photochemical model

As noted in the Methods, previously published models of the Venusian atmosphere did not include the scavenging effect of $PH_3$, and so we found it necessary to develop our own model. Additional publications are planned on this work, expanding on the description below.

We employ the 1D photochemistry-diffusion code ARGO[46] to solve the atmospheric transport equation for the steady-state vertical composition profile. The model took the handful of known reactions between $PH_3$ and the major reactive Venusian species O, Cl, OH, and H, and combined them with the previously published Venus atmospheric networks of Krasnopolsky[49,57] and Zhang[50], and the network of Rimmer & Rugheimer[48]. This whole-atmosphere model allows us to self-consistently assess the lifetime of $PH_3$ throughout the atmosphere. The model accounts for photochemistry, thermochemistry and chemical diffusion.

We employ a 1D photochemistry-diffusion code, called ARGO[46], to solve for atmospheric transport. ARGO is a Lagrangian photochemistry/diffusion code that follows a single parcel as it moves from the bottom to the top of the atmosphere, determined by a prescribed temperature profile. The temperature, pressure, and actinic ultraviolet flux are updated at each height in the atmosphere. In this reference frame, bulk diffusion terms are accounted for by time-dependence of the chemical production, $P_i$ (cm$^3$ s$^{-1}$), and loss, $L_i$ (s$^{-1}$), and so below the homopause, the chemical equation being solved is effectively:

$$\frac{\partial n_i}{\partial t} = P_i[t(z, v_v)] - L_i[t(z, v_z)]n_i, \qquad M1$$

where $n_i$ (cm$^{-3}$) is the number density of species $i$, $t$ (s) is time, $z$ [cm] is atmospheric height, and $v_z = K_{zz}/H_0$ (cm/s) is the effective vertical velocity due to Eddy diffusion, from the Eddy diffusion coefficient $K_{zz}$(cm$^2$ s$^{-1}$). The model is run until every major and significant minor species (any with $n_i > 10^5$ cm$^{-3}$) agrees between two global iterations to within 1%. We have modified the UV transport calculation in two ways. First, we ignore the absorption of $SO_2$ for the first three global iterations, and include it afterwards. This seems to help the model to converge. In addition, we have included a 'mysterious absorber' with properties[47]:

$$\frac{d\tau}{dz} = 0.056/km\, e^{-(z-67\text{ km})/3\text{ km}} e^{-(\lambda-3600\text{ Å})/1000\text{ Å}}, \qquad z > 67\text{ km};$$

$$\frac{d\tau}{dz} = 0.056/km\, e^{-(\lambda-3600\text{ Å})/1000\text{ Å}}, \qquad 58\, km \le z \le 67\text{ km};$$

$$\frac{d\tau}{dz} = 0, \qquad z \le 58\text{ km};$$

With these conditions, using the photochemical network described below, it took 33 global iterations for the calculations to converge.

***Photochemical network for Venus***

For our chemical network, we use STAND2019[48], which includes H/C/N/O species. We have also added a limited S/Cl/P network relevant for the Venusian atmosphere by copying the low atmospheric network of Krasnopolsky[47,49] and the middle atmosphere network of Zhang[50], and supplementing those networks with the following reactions (see ref. 10 for literature sources):

$$PH_3 + M \rightarrow PH_2 + H + M, \qquad k_0 = 3.4 \times 10^{-8}\text{ cm}^3\text{ s}^{-1}\text{ e}^{-35644\text{ K/T}};$$

$$k_\infty = 1.91 \times 10^{18}\text{ s}^{-1}\text{ e}^{-40063\text{ K/T}};$$

$$PH_3 + H \rightarrow PH_2 + H_2, \qquad k = 7.22 \times 10^{-11}\text{ cm}^3\text{s}^{-1} e^{-884\text{ K}/T};$$

$$PH_3 + O \rightarrow \text{Products}, \qquad k = 4.75 \times 10^{-11}\text{ cm}^3\text{ s}^{-1};$$

$$PH_3 + OH \rightarrow PH_2 + H_2O, \qquad k = 1.61 \times 10^{-11}\text{ cm}^3\text{ s}^{-1};$$

$$PH_3 + Cl \rightarrow PH_2 + HCl, \qquad k = 2.4 \times 10^{-10}\text{ cm}^3\text{ s}^{-1};$$

$$PH_3 + h\nu \rightarrow PH_2 + H$$

We calculate the rate constant for $PH_3 + M \rightarrow PH_2 + H + M$, where M is any molecule, as described below. For the reaction, $PH_3 + O \rightarrow$ Products has products with branching ratios (when T < 500 K) of >90% $PH_2O$ + H and <10% $PH_2$ + OH. Because virtually no $PH_2$ is restored to $PH_3$ in the photochemical model, we do not explicitly include $PH_2O$ in the chemical network, but simply treat the reaction of $PH_3$ + O as a loss of $PH_3$ with the above given rate constant.

For the reactions we include not involving $PH_3$, we use the networks of Krasnopolsky[47] and Zhang[50] with the following modifications. The network of Krasnopolsky include prescribed reverse reactions. We neglected these, including the forward reactions and the thermochemical constants from Burcat[51] for calculating the reverse reactions for those species already included in STAND2019, as well as reactions that include the species S, $S_2$, $S_3$, $S_4$, $S_5$, $S_6$, $S_7$, $S_8$, HS, SO, ClO, ClS, $Cl_2$, $H_2S$, OCS, $SO_2$, $SO_3$, $S_2O$, HOCl, ClCO, $Cl_2S$, $Cl_2S_2$, $HSO_3$, $H_2SO_4$, $PH_2$ and $PH_3$, in the manner described by refs. 46,52. We added reverse reactions for the reactions from the Zhang[50] middle atmosphere network wherever possible. In addition to these reactions, we include condensation of $S_n$ species[53] and sulfuric acid $H_2SO_4$[54].

We add removal of $SO_2$ into the clouds in order to match the top boundary conditions from the lower atmosphere models[47] to the bottom boundary conditions for the middle atmosphere models, where the $SO_2$ is depleted by orders of magnitude[50]. Bierson and Zhang accomplished this by depleting $SO_2$ via oxidation to $SO_3$ and removal by reaction with $H_2O$ to form sulfuric acid[55]. The $SO_2$ could only be brought into agreement with observation by decreasing $K_{zz}$ within the cloud layer and fixing $H_2O$ to be equal to observed concentrations throughout the atmosphere. We do not fix the $H_2O$ concentrations, and so instead have depleted $SO_2$ by including rainout with a Henry's Law approximation modified as described by Sander (ref. 56, their Section 2.7), with the pKa of the resulting sulfurous acid being 1.81 and the pH of Venus's clouds as -1.5. It should be noted that the Henry's Law constants from Sander[56] are for liquid water, and not for sulfuric acid droplets. Nevertheless, incorporating this loss term brings the $SO_2$ curve into better agreement with observation, and may instead be interpreted as approximating photochemical loss of $SO_2$ via a different mechanism or series of reactions.

### *Thermal Decomposition of Phosphine*

Consideration of the thermal decomposition of phosphine is important because concentrations of radicals below the clouds of Venus are trace, << 1 ppb. Estimated concentrations of radicals below the clouds are very uncertain, and even with the largest published predictions[47] for radical concentrations in the lower atmosphere of ~1000 $cm^{-3}$, thermal decomposition dominates $PH_3$ destruction and therefore is determinative of the lifetime near the surface of Venus. The thermal decomposition of $PH_3$ has been considered theoretically[65]. Theoretical values of $k_{\mathrm{uni}}$ ($s^{-1}$) and $k_\infty$ ($s^{-1}$) are given as[65]:

$$k_{\mathrm{uni}} = 3.55 \times 10^{14}\ \mathrm{s}^{-1} e^{-35644\ \mathrm{K}/T}$$

$$k_\infty = 1.91 \times 10^{18}\ \mathrm{s}^{-1} e^{-40063\ \mathrm{K}/T}$$

but no value for the rate constant at the low-pressure limit, $k_0$ ($cm^3\ s^{-1}$), is given. This rate constant needs to be determined in order to calculate the rate constant over a wide range of pressures using the Lindemann expression:

$$k = \frac{k_\infty}{1 + k_\infty/(k_0[M])} \qquad R3$$

where [M] is the number density of the third body, in our case $[\mathrm{M}] = n$, where $n$ ($cm^{-3}$) is the atmospheric number density. The rate constant at the low pressure limit can be estimated by considering that $k_{\mathrm{uni}}$ was calculated for 1300 bar and 900 K, so $[\mathrm{M}] = 1.07 \times 10^{22}\mathrm{cm}^{-3}$, and solving Equation (*R3*) with $k = k_{\mathrm{uni}}$. Doing so yields:

$$k_0 = 3.4 \times 10^{-8}\ \mathrm{cm}^3\ \mathrm{s}^{-1} e^{-35644\ \mathrm{K}/T} \qquad R4$$

An alternative way to estimate $k_0$ from $k_\infty$ is to perform a simple conversion of units, with $k_0 = kT/(1\ bar)\ k_\infty$, which gives:

$$k_0 = 2.6 \times 10^{-4}\ \mathrm{cm}^3\ \mathrm{s}^{-1} \left(\frac{T}{300\ K}\right) e^{-40063\ \mathrm{K}/T} \qquad R5$$

Finally, we can consider the decomposition of $NH_3$ as an analogue of the decomposition of $PH_3$. In this case, the low-pressure limit for $NH_3$ has been experimentally determined over a temperature range of 1740-3300 K[66]. We assume that the prefactor is the same for $PH_3$ and that the only difference is the activation energy, for which we compare the activation energy at the high pressure limit of 40063 K for $PH_3$[65] to the activation energy at the high-pressure limit of 48840 K for $NH_3$[65], and multiply the ratio of these activation energies to the measured activation energy for $NH_3$ at the low-pressure limit of 39960 K[66] to find:

$$k_0 = 7.2 \times 10^{-9} \text{ cm}^3 \text{ s}^{-1} \, e^{-32778 \text{ K}/T} \qquad R6$$

The timescales for thermal decomposition derived from these rate constants, along with the timescale using only $k_\infty$, are plotted in Supplementary Figure 8. Since our first estimate, Equation (*R4*), yields the longest timescale, and will therefore be most favorable for abiotic $PH_3$ scenarios, we use that value.

In total, the network includes 460 species and 3406 forward reactions, including 173 photochemical reactions. For a full description of the photochemical network models that we used in this paper, see ref. 35.

### *Initial and boundary conditions*

In modelling the Venusian atmosphere, we follow refs. 47,57 in taking the temperature-pressure (TP) profile from the Venus International Reference Atmosphere (VIRA). Specifically, we use previously published TP profiles of ref. 58 for the deep atmosphere profile (0-32 km) and for the altitudes between 32-100 km, where we use the 45 degrees latitude profile. For the altitudes between 100-112 km we use the VIRA dayside profile from ref. 59. Supplementary Figure 8 shows the temperature-pressure profile adopted in this work. We similarly follow refs. 47,57 in the Eddy diffusion profile, taking it to be constant at $2.2 \cdot 10^3$ $cm^2$ $s^{-1}$ for z <30 km, $1 \cdot 10^4$ $cm^2$ $s^{-1}$ for z = 47-60 km, $1 \cdot 10^7$ $cm^2$ $s^{-1}$ for z >100 km, and connected exponentially at intermediate altitudes. Supplementary Figure 8 shows the Eddy diffusion profile adopted in this work (see also below, under Photochemistry in the atmosphere).

We take fixed surface boundary conditions from ref. 47 for the major atmospheric species, and with initial surface boundary conditions from ref. 47 for minor species, radicals and atoms, except for $PH_3$. Initial surface abundances for our model are shown in Supplementary Table 2.

We include a source of $PH_3$ in the clouds, with flux:

$$\Phi(z) = 0.5\Phi_0 \, [\tanh\left(\frac{z - 45 \, km}{2 \, km}\right) \tanh\left(\frac{65 \, km - z}{2 \, km}\right) + 1]$$

where $\Phi(z)$ ($cm^{-2}$ $s^{-1}$) is the $PH_3$ flux at height z (km), and $\Phi_0 = 10^7$ ($cm^{-2}$ $s^{-1}$) is assigned to reproduce the 10 ppb $PH_3$ concentrations.

### *Validation and limits of the model: Comparison of the Venus chemical kinetics model with atmospheric observations*

The model predictions for a variety of species, CO, $O_2$, OCS, $H_2O$, $SO_2$, $H_2S$, HCl, $S_3$, SO, and $PH_3$, compared to observations (Supplementary Table 3) are plotted in Supplementary Figure 9. The profiles for all species agree with observations within an order of magnitude in concentration, within 5 km height, except for the ALMA $SO_2$ data (ref. 27; this work), water vapor and $O_2$. Photolysis of water is very efficient for our model, and depletion by reaction with $SO_3$ is significant, and our model's predicted water vapor drops off rapidly above 70 km, leading to a discrepency between observed $H_2O$ and model $H_2O$ of several orders of magnitude. This discrepancy is accompanied by higher concentrations of OH and O above 70 km, and so we probably underestimate the lifetime for phosphine above 70 km. However, the lifetime of phosphine at these heights is very short, on the order of days to seconds, for all published models of Venus's middle atmosphere (e.g. by Zhang[50] and Bierson & Zhang[55]). Our model also predicts too much $O_2$ in the middle atmosphere of Venus, with similar concentrations as Zhang[50] and Bierson & Zhang[55.]

We consider the possibility that our model contains an idiosyncrasy or error that leads to significant underestimates of $PH_3$ lifetime, and hence overestimates the difficulty of abiotic buildup. To assess this possibility, we repeat our calculations of $PH_3$ lifetime and required production rates using concentration profiles of H, OH, O, Cl, and $SO_2$ drawn from Bierson & Zhang[55] (aided by C. Bierson, personal communication, 08/02/2019). This model excludes $PH_3$; consequently, it may overestimate lower-atmosphere radical abundances and underestimate $PH_3$ lifetimes. Use of these radical profiles, instead of the profiles drawn from our model, result in $PH_3$ lifetimes becoming short ($<10^3$ s) at an altitude of 71 or 80 km instead of 63 km in our model, depending on which of the scenarios from Bierson & Zhang we adopt (their nominal vs. their low $K_{zz}$+$S_8$ scenarios). However, this change in destruction altitude does not affect the upper limits on lifetime we calculate strongly enough to affect the conclusions of this paper.

The full model output is included in the additional materials. For further description of the models used in this paper also see ref. 35.

## Potential pathways for phosphine production.

We discuss the potential pathways for phosphine production in the Venusian environment and why $PH_3$ production is ruled out for conditions of the Venusian atmosphere, surface, and subsurface. The discussion is summarized in Table S4. For more details, see ref. 35.

### *Phosphine lifetime*

We estimate the rate of destruction of phosphine as discussed above.

The photochemical lifetime of $PH_3$ can be high in the deep atmosphere, but is always low in the high atmosphere where UV photolysis and its concomitant radicals efficiently destroy $PH_3$. In the deep atmosphere, transport to the upper atmosphere will limit $PH_3$ lifetime. To account for the effects of transport on $PH_3$ lifetime, we calculate the transport timescale for $PH_3$ at altitude $z_1$ due to eddy diffusion, via $t_{transport}=\Delta z^2/K_{zz}$, where $K_{zz}$ is the eddy diffusion coefficient, and $\Delta z=z_0-z_1$, where $z_0$ is the vertical altitude at which $PH_3$ lifetimes are short due to photochemistry. In our photochemical model above, $z_0$ = 63 km; however, in other models, $z_0$ can be as high as 80 km (C. Bierson, personal communication, 07/24/2019). The pseudo-first order rate constant of $PH_3$ loss due to eddy diffusion is thus $1/ t_{transport}= K_{zz} /\Delta z^2$. We conservatively adopt $K_{zz}=K_{zz}(z_1)$; since Kzz is monotonically nondecreasing with z, this underestimates $K_{zz}$, overestimates $t_{transport}$, and underestimates the destruction rate. We adopt the lower of the transport timescale to 63 km and the photochemical lifetime as our overall lifetime.

### *Photochemical production of phosphine*

The maximum possible potential rate at which phosphine could be produced photochemically is estimated as follows. We created a network of reactions for which kinetic parameters are known[67] that could lead from $H_3PO_4$ (phosphoric acid) to $PH_3$ (phosphine), by reaction with radical species in the Venusian atmosphere (Supplementary Figure 7). We know from the thermodynamic data summarized below (and presented in detail in ref. 35) that reaction with stable species such as $H_2$ cannot yield phosphine in adequate amounts from thermodynamic arguments. Where reactions were possible but no kinetic data for the phosphorus species was known, homologous nitrogen species reaction kinetics were used instead. This was validated by comparing reaction of analogous nitrogen and phosphorus species: the N=O and P=O bond energies are similar, transition state energies for cleavage of H-N=O and H-N=P are similar, HOMO/LUMO structures (which relate to reaction pathways) are similar, and $HNO_3$ and $HNO_2$ have similar radical reaction kinetics to $HPO_3$ and $HPO_2$ species respectively. By contrast, hydrogen abstraction from $NH_3$ is very much slower than hydrogen abstraction from $PH_3$, reflecting different bond strengths; however kinetic data for P species is known for hydrogen abstraction reactions. Kinetic data were obtained from the NIST kinetics database[60], supplemented by refs. 67 and 68.

The kinetics of each reaction was calculated for densities of radical species calculated in the photochemical code above. The *maximum possible* rate of phosphine production was then calculated with the following assumptions:

- that there were no 'back' reactions (i.e. all reactions that could happen were ones that reduced phosphorus atoms)
- for *each* reaction, the concentration of phosphorus species taking part was the concentration of *all* phosphorus species in the atmosphere.

The maximum rate through the network is therefore the maximum rate through any path through the network, and the rate through any path is the minimum of each individual reaction rate along that path. This maximum possible rate was compared with the rate needed to balance photochemical destruction of phosphine. The steady-state abundance in this case is:

$$f(PH_3) = \frac{P_{abiotic}(PH_3)}{L(PH_3)n}$$

where $P_{abiotic}(PH_3)$ ($cm^{-3}$ $s^{-1}$) is the maximum abiotic production rate of $PH_3$, n ($cm^{-3}$) is the total gas number density, and

$$L(PH_3) = \max\{L_R, 1/\tau_{transport}\}$$

is the $PH_3$ destruction rate constant ($s^{-1}$). Here, $L_R$ is the destruction rate of $PH_3$ according to our photochemical model, and $1/\tau_{transport} = K_{zz}/\Delta z^2$, as defined above.

### *Equilibrium thermodynamics in the atmosphere and surface*

Chemical thermodynamics is a robust way to determine if a reaction favors the net production of the product at equilibrium (i.e. does not require an input of energy from the environment to drive the net conversion of reagents to product). A thermodynamically possible reaction may be prohibited by kinetic barriers, but a thermodynamically impossible reaction will not be a spontaneous net producer of a reaction product no matter what the reaction kinetics. In other words, while thermodynamics always correctly predicts whether a given process is favored (is spontaneous from the thermodynamic point of view), it does not tell if the chemical process will take place at an observable rate. (Apropos the kinetic model above, we note that the reaction of hydrogen *atoms* with $H_3PO_4$ to form $PH_3$ is thermodynamically favoured.)

Considering the atmosphere and surface of Venus, we created a list of chemicals, their concentrations, and reactions, for all potential $PH_3$ production pathways. To produce phosphine from the oxidized phosphorus species expected on Venus, a phosphorus-containing compound must be reduced. We therefore considered examples of all types of phosphorous-containing compounds as sources of phosphorus, and as reducing agents all atmospheric components which could in principle behave as reducing agents for which an atmospheric concentration has been measured or modeled, and a sample of Fe(II) solid compounds which could act as surface reducing agents. We used literature values for standard free energy and the concentrations of reactants. Temperatures and pressures come from the Venusian Standard Atmosphere. Activities were calculated using Berthelot's equation[69] and literature values for critical temperature and pressure for gases. We have calculated Gibbs Free Energy[70,71] of reaction to estimate the possibility of phosphine formation (Supplementary Figure 7), recalling that negative values mean the reaction happens spontaneously. Reactions of $P_4O_6$, $P_4O_{10}$, $H_3PO_4$ and $H_3PO_3$ were considered (the last of these only in solution phase in the clouds), as well as surface reduction of phosphate minerals. Our calculation considered 8 phosphorus-reducing species, in ~75 reactions. Reactions with gases were calculated with a high or a low gas concentration, derived from published models[47,57,72-76], in all combinations., in different temperature/pressure regimes, leading to ~3840 conditions tested for each reaction. Thermodynamics was only followed to the cloud tops, after which freeze-out of phosphorus species makes reactions of stable phosphorus compounds implausible.

Not surprisingly, for an oxidized atmosphere with little P, none of the reactions favor the formation of phosphine, on average having a free energy of reaction of +100kJ/mol (Supplementary Figure 7). For further details on thermodynamic modeling of phosphine production in the Venusian atmosphere see ref. 35 .

As an example of our approach, we present a calculation for phosphorous acid ($H_3PO_3$). This compound will spontaneously decompose on heating to form phosphoric acid and phosphine; this is a standard laboratory method for making phosphine . Phosphorous acid is not stable in gas phase, but could in principle be formed in cloud droplets by reduction of phosphoric acid. We calculated the equilibrium fraction of phosphorus present as $H_3PO_3$ in droplets in Venus' cloud assuming conditions as described above. $H_3PO_3$ is a tiny fraction of the total phosphorus inventory; if phosphorus was present as 1 molar $H_3PO_4$ in droplets, then the concentration of $H_3PO_3$ would be ~$6.10^{-17}$ molar at 47 km (the base of the clouds, which is below our minimum detection altitude), and at >53 km (the approximate minimum altitude of the candidate detection of $PH_3$) would be ~$10^{-20}$ molar. If we assume a total volume of cloud material of $1.04 \cdot 10^{10}$ $m^3$ (calculated from the droplet sizes, droplet abundances and cloud depth), the clouds above 47 km would contain ~44 milligrams of $H_3PO_3$ in the entire Venusian atmosphere. Reduction of phosphate to phosphite and subsequent disproportionation of phosphite is therefore not a plausible source of phosphine.

We investigated possible chemical reactions involving subsurface phosphate-containing minerals that may produce phosphine by reduction. The chemical composition of the subsurface and the deep interior of Venus is poorly known. It is assumed to be similar to the chemical composition of the Earth's crust and mantle, mainly due to the similarity of Earth and Venus in terms of the size and the overall bulk density of two planets[77]. The Venusian crust is largely basalt, which suggests that Venusian mantle is likely similar in chemical composition to Earth's mantle[75,77,78].

Detailed modelling of the subsurface chemistry is not practical. A convenient simplification of this complexity with respect to the redox state of the crust, and hence the potential for the crust to support redox reactions, is the concept of oxygen fugacity ($fO_2$)[61]. In brief, $f(O_2)$ is the notional concentration of free oxygen in the crustal rocks. A higher oxygen fugacity of the rock means a more oxidized rock and a lower probability of reduction of phosphates.

We use the concept of oxygen fugacity[61] to estimate the likelihood of phosphine production from phosphate subsurface minerals. We model the equilibrium between phosphate and phosphine under oxygen fugacity buffers QIF, WM, IM, FMQ or MH (terrestrial crustal rocks typically have $f(O_2)$ between FMQ and MH). Calculations were performed for temperatures between 700K and 1800K, at 100 or 1000 bar and with 0.01%, 0.2% and 5% water. We conclude that oxygen fugacity of plausible crust and mantle rocks is 8-15 orders of magnitude too high to support reduction of phosphate. Degassing of mantle rocks would therefore produce only trivial amounts of phosphine compared to their total phosphorus content.

Our fugacity calculations are supported by observations that phosphine is not known to be produced by volcanoes on Earth. Calculations on the production of $PH_3$ through volcanism on a simulated anoxic early Earth showed that only trace amounts of volcanic $PH_3$ can be created through this avenue; the predicted maximum production rate is only 102 tons per year[79]. We note that the estimation of the maximum production of $PH_3$ through the volcanic processes reported by ref. 79 is made under the assumption of a highly reduced planet, which provides favorable conditions for $PH_3$ volcanic production. The volcanic production of $PH_3$ in more oxidized planetary scenarios is even more unlikely.

### *Lightning*

Lightning may be capable of producing a plethora of molecules that at first glance are energetically unfavorable to make.

We assume for simplicity that the energy delivered by a lightning bolt leads to a complete atomization of chemicals within a droplet or within the gas inside the volume of the lightning stroke. We further assume that subsequent recombination of atoms into stable chemicals proceeds at random, and is dependent solely on the composition of Venus' clouds and atmosphere and the number density of the atoms in the resulting plasma.

Under such assumptions, the maximum amount of phosphine produced in 1 year is $3.38 \cdot 10^8$ grams. If this accumulated in Venus' atmosphere for a full Venusian year without any destruction, it would reach a partial pressure of 0.00076 parts per trillion, much lower than ~10 ppb phosphine concentration that we model in the Venusian atmosphere.

The above calculations are in agreement with several studies on the formation of reduced phosphorus species, including $PH_3$, as a result of simulated lightning discharges in laboratory conditions. The laboratory experiments suggest that in principle reduction of phosphate to phosphine through lightning strikes is extremely inefficient[80,81], further strengthening the argument that lightning discharges cannot be responsible for the observed phosphine concentrations in the Venusian atmosphere.

### *Meteoritic delivery*

Iron-nickel meteorites are known to contain reduced species of phosphorus. We calculate possible delivery fluxes on Venus to exclude phosphine production by meteoritic delivery.

The current accretion rate of meteoritic material to the Earth is of the order of 20-70 kilotonnes/year[82]. Approximately 6% of this material is in the form of phosphide-containing iron/nickel meteorites[83] which contain an average of 0.25% phosphorus by weight[84]. If we assume that hydrolysis of $(Fe,Ni)_3P$ phosphides to phosphine is 100% efficient, that would deliver a maximum of ~10 tonnes of phosphine to the Earth every year, or about 110 milligrams/second. Assuming Venus accretes phosphides at a similar rate, and assuming a photochemical destruction rate of phosphine is as shown in Figure 5, and that all the phosphine is deposited at 50-60 km (where destruction is slowest), the concentrations of $PH_3$ is:

$$\mathrm{f(PH_3)} = \frac{\mathrm{N_A S_M \tau(PH_3)}}{\mathrm{\mu(PH_3)\ n\Delta h\ 4\pi R_{Venus}^2}} \qquad R1$$

where $\mathrm{\mu(PH_3)} = 34$ g/mol is the molar mass of $PH_3$, $\mathrm{n} = 6.5 \times 10^{18}$ $\mathrm{cm^{-3}}$ is the gas density at 60 km, $\Delta\mathrm{h} = 10$ km, $\mathrm{R_{Venus}} = 6052$ km is the radius of Venus, $\mathrm{N_A} = 6.022 \times 10^{23}$ is Avagadro's constant, $\tau(\mathrm{PH_3}) = 10^{10}$ s and $\mathrm{S_M} = 0.11$ g/s is the amount of $PH_3$ delivered per second to Venus according to the above approximation. Applying all these values, we find:

$$\mathrm{f(PH_3)} = 6.5 \times 10^{-13} \qquad R2$$

This is orders of magnitude lower than the observationally-constrained ~20 ppb phosphine concentrations, regardless of the photochemical model in question. We note that the above calculations are also in agreement with previous estimations of the phosphine production through meteoritic delivery, which were also found to be negligible[84].

Other endergonic processes are covered in ref. 35.

### *Horizontal Transport, Chemical Timescales and Latitudinal Variation of Phosphine*

As discussed above, the lifetime of $PH_3$ in the atmosphere of Venus is constrained by photodissociation in the upper atmosphere, thermal decomposition near the surface, and diffusion in between. The constraints these three processes provide are similar across all the photochemical models. Further constraints involving radical chemistry are highly model dependent. We can use the latitudinal variation of $PH_3$ to constrain the lifetimes of $PH_3$ within the atmospheric regions where radical chemistry dominates. We do this by comparing the model-dependent chemical lifetimes of $PH_3$ to zonal and meridional velocities. If the velocities are too great, compared to chemical lifetimes, mixing will be efficient, and regardless of the source of $PH_3$, it will be well-mixed. The minimum scale over which variation is expected is then:

$$L = v\tau_{\mathrm{chem}} \qquad \mathrm{R7}$$

where $v$ (m/s) is the either the meridional or zonal velocity in question and $\tau_{\mathrm{chem}}$ (s) is the chemical lifetime of $PH_3$. We consider that observable latitudinal variation implies that $L \lesssim R_{\mathrm{Venus}}$, and plot the velocity above which $PH_3$ would be well-mixed, in Supplementary Figure 9. Given the constraints on zonal and meridional velocity given by cloud tracers[63,64], the latitudinal variation is somewhat unexpected. It implies either that (a) the $PH_3$ observed is largely above 60 km *and* that this work *and* the models presented in this work and in refs 47,49 correctly predict or under-predict the above-cloud concentrations of OH, H, O and Cl, or (b) that there is some unknown mechanism that more rapidly destroys $PH_3$ within the cloud layer, or both. Further observations will help constrain both the robustness of the latitudinal variation and its vertical profile. In addition, further laboratory experiments exploring $PH_3$ stability will help to identify what other mechanisms may be responsible for its destruction within the cloud layer.

## $PH_3$ and hypotheses on Venusian life

The Venusian surface is widely believed to be uninhabitable. However, the clouds of Venus offer temperate conditions, and the possibility of life in the Venusian clouds has been discussed for decades[16,85-87]. It has been speculated that the unknown and variable ultraviolet opacity (currently being monitored by the *Akatsuki* spacecraft) is due to life particle pigments[17] though chemical processes may be the source[88].

To this discussion, we add that trace $PH_3$ in Earth's atmosphere is uniquely associated with biological and anthropogenic activity[10,18,89], and we have proposed that any detectable phosphine found in the atmosphere of a rocky planet is a promising sign of life[10]. Terrestrial life produces this highly reducing, endothermic gas even in an overall oxidizing surface environment, and notably the biological production of phosphine is found to be favored by cool, acid conditions[18]. Thus, phosphine is a biosignature of some interest for the cool but hyper-acidic conditions of Venusian clouds.

Computer models of Venus[90,91] have shown that a habitable surface with liquid water could have persisted up to 715 million years ago (Mya), and *Magellan* data indicate complete resurfacing after this (~500 ±200 Mya[92], perhaps taking ~100 Myr[93]). Such effects would radically alter the planetary environment, and it has been widely suggested that, as the surface was rendered inhospitable, life could have gradually colonized the clouds, eventually becoming an entirely aerial biosphere. As a parallel for such suggestions, Earth has a metabolically-active aerial biosphere[94,95].

Initial modelling based on terrestrial biochemistry suggests that biochemical reduction of phosphate to phosphine is thermodynamically feasible under Venus cloud conditions[35]. Biological phosphine production on Venus is likely to be energy requiring[18]. However, life can make substantial energy investment into compounds that provide important biological functionality. There are many potential useful biological functions including signaling, defense, or metal capture for which phosphine has useful properties[89], so endergonic biosynthesis cannot be ruled out.

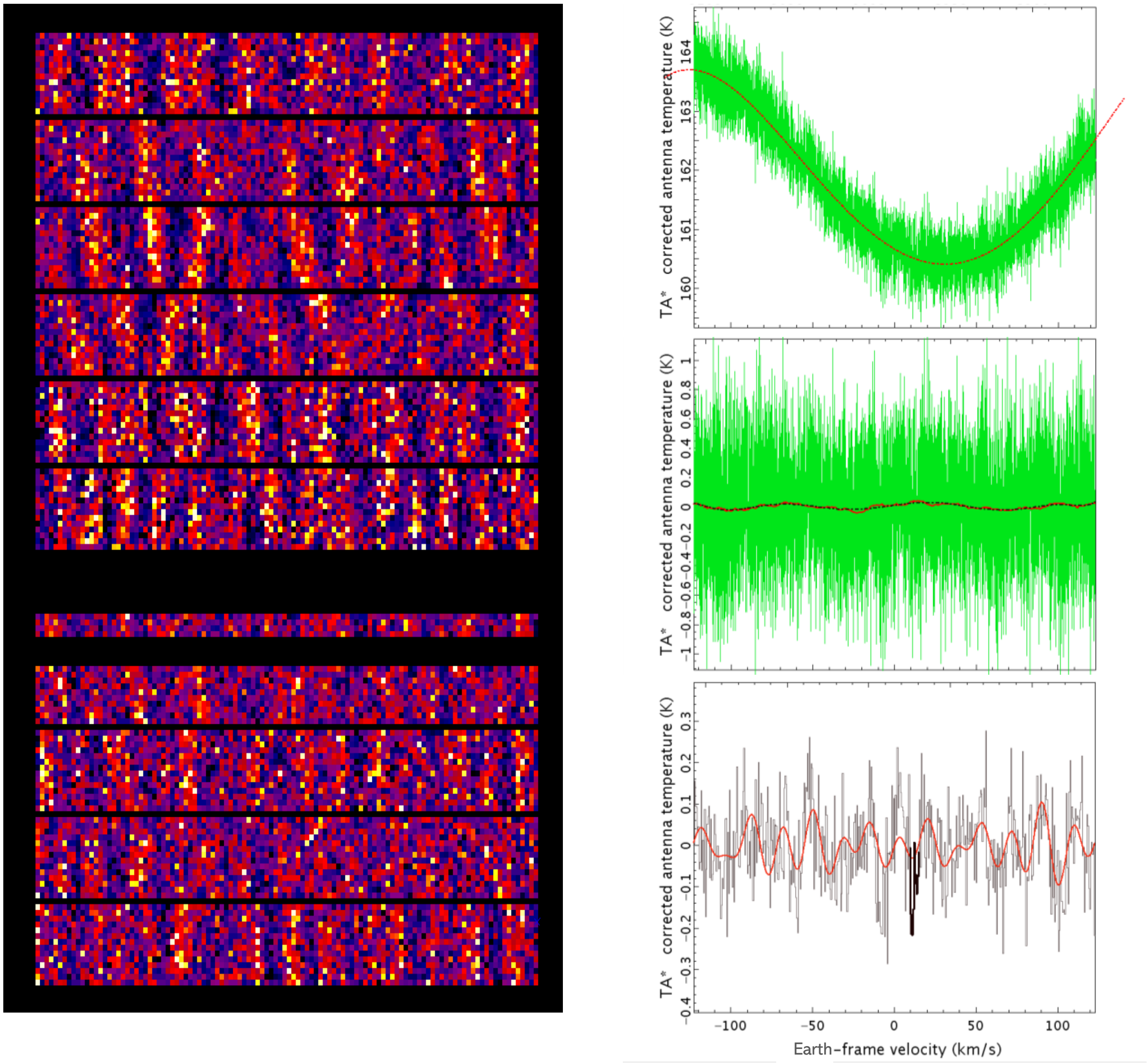

**Supplementary Figure 1.** Instrumental effects present in the JCMT data are illustrated. Data are shown after an initial step of blanking 512 channels with increasing noise towards each end of the 8192 channel passband. Left: the 140 sub-observations are stacked vertically, with spectral channel on the X-axis and time on the Y-axis (earliest observation at the bottom). The black bars every $15^{th}$ row denote gaps between observations. Signatures of reflected signals have here been fitted and subtracted, leaving the ripples with ~8,16 periods across 250 MHz. The spectra have been binned to 2.2 km/s velocity resolution for clarity; the Doppler-shifted Venus absorption is then centered around channel 62 of 112, counting from the left. Right: stages of the reduction for observation 1 (bottom 14 rows of data in the left panel; note that this example is for demonstration, and script Data S1 in fact reduces every row in the left panel separately). Top-right panel shows the $4^{th}$-order polynomial fit (red dotted curve) to the full passband of observation 1. Middle-right panel shows the subsequent residual, overlaid with a median-filter (red curve) and the $9^{th}$-order fit to this filtered data (black dashed curve). Bottom-right panel shows the next subsequent residual, with the data binned to 0.55 km/s resolution (a section around Venus' velocity is highlighted with heavier bars). The overlaid red curve demonstrates the trend derived from the main Fourier components identified in the spectral ripple (via kappa task 'fourier').

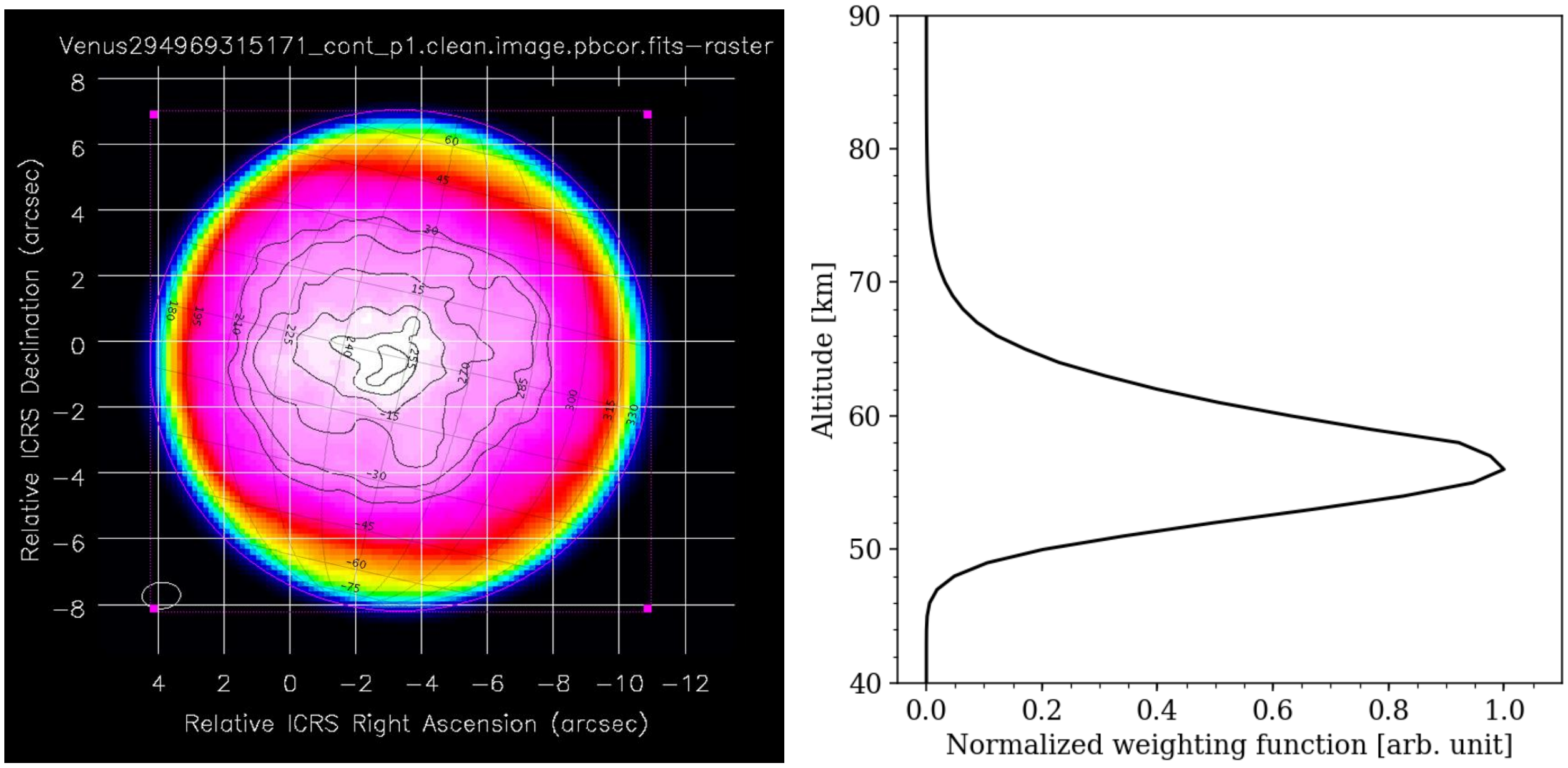

**Supplementary Figure 2.** Geometry of Venus is presented on the sky and by altitude. Left panel: Illustration of the orientation of Venus as viewed at the time of our ALMA observation. The sub-Earth point (center) is at longitude $256^o$ and latitude $-0.6^o$. The Sun was overhead at longitude $194^o$ and latitude $+0.2^o$, hence longitudes beyond the terminator (> $284^o$) were in darkness. (For comparison, during JCMT observations, the planet was just over half illuminated, with the sub-solar point closer to the left limb.) Planetary rotation is from right to left. The color scale shows the continuum signal in our observations, illustrating that the polar caps appear cooler. The overlaid contours were only used for checking alignment of the longitude/latitude grid, and do not show real structures (contour spacings are of order the noise of ~0.1 Jy/beam; this is higher than the spectral channel noise due to dynamic range limitations with all baselines included). Magenta outlines were also temporary guides. The ellipse at lower-left indicates the size and orientation of the ALMA beam for the continuum data (the beam for the line data is very similar). Right panel: Illustration of the altitude-range above which the phosphine absorption can originate. The weighting function shows the altitudes where the continuum (thermal) emission arises, at 266 GHz (near the $PH_3$ 1-0 frequency but not affected by the absorption). The function peaks at 56 km and its FWHM spans approximately 53 to 61 km. The effect of uncertainties in the temperature profile of the Venusian atmosphere is to introduce systematics of order 2-3 km. The continuum emission has very high opacity, so our absorption observations do not trace altitudes below the peak of the weighting function.

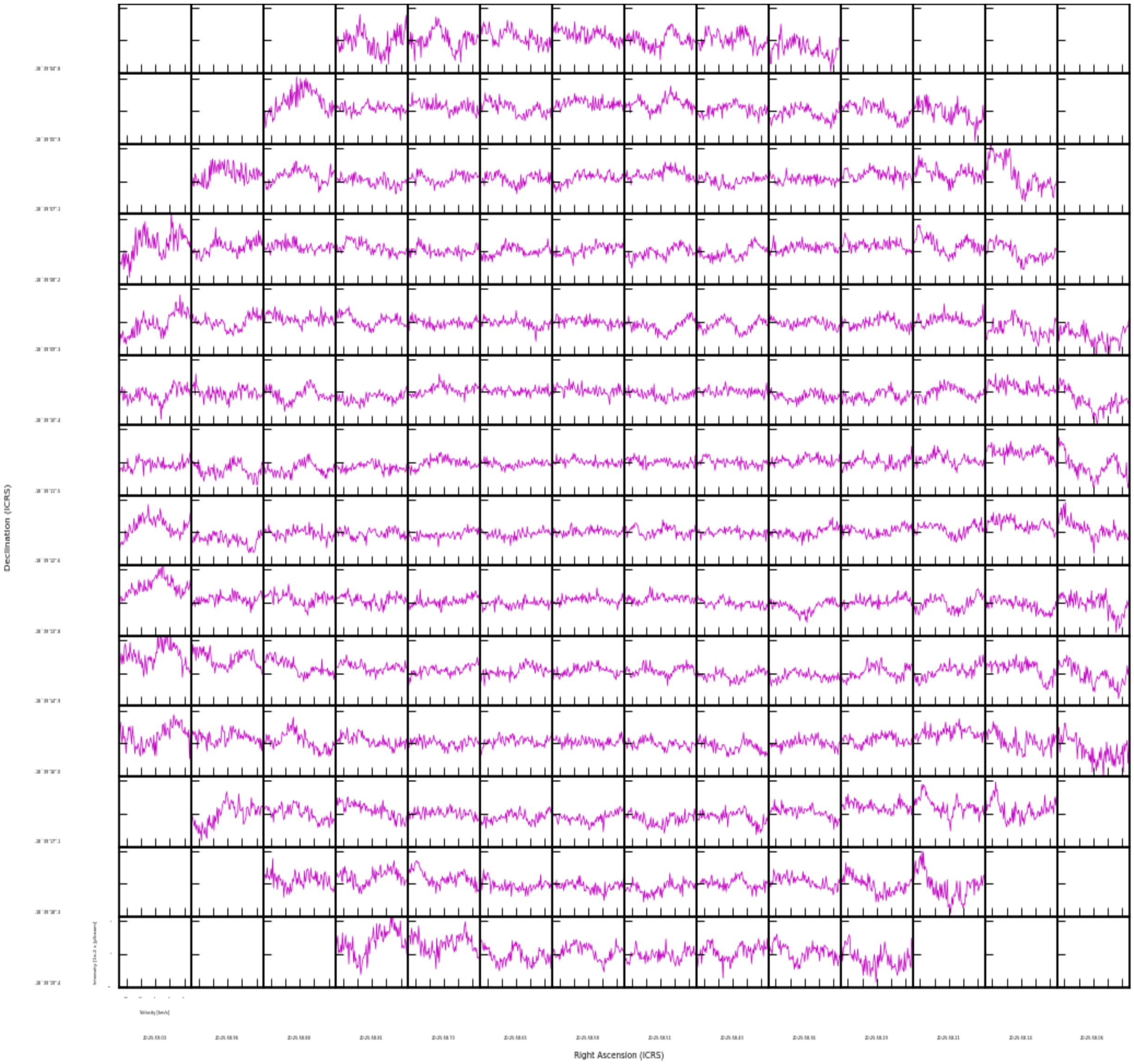

**Supplementary Figure 3.** Grid of $PH_3$ 1-0 spectra from ALMA is presented, illustrating the difficulties of detecting the phosphine line on scales of the restoring beam. Each sub-plot spans 1.1 arcseconds on the planet (which has the same orientation as in Supplementary Figure 2) and has an X-axis velocity range of ~ ±25 km/s. Blank boxes lie outside the planet (image mask has been applied).

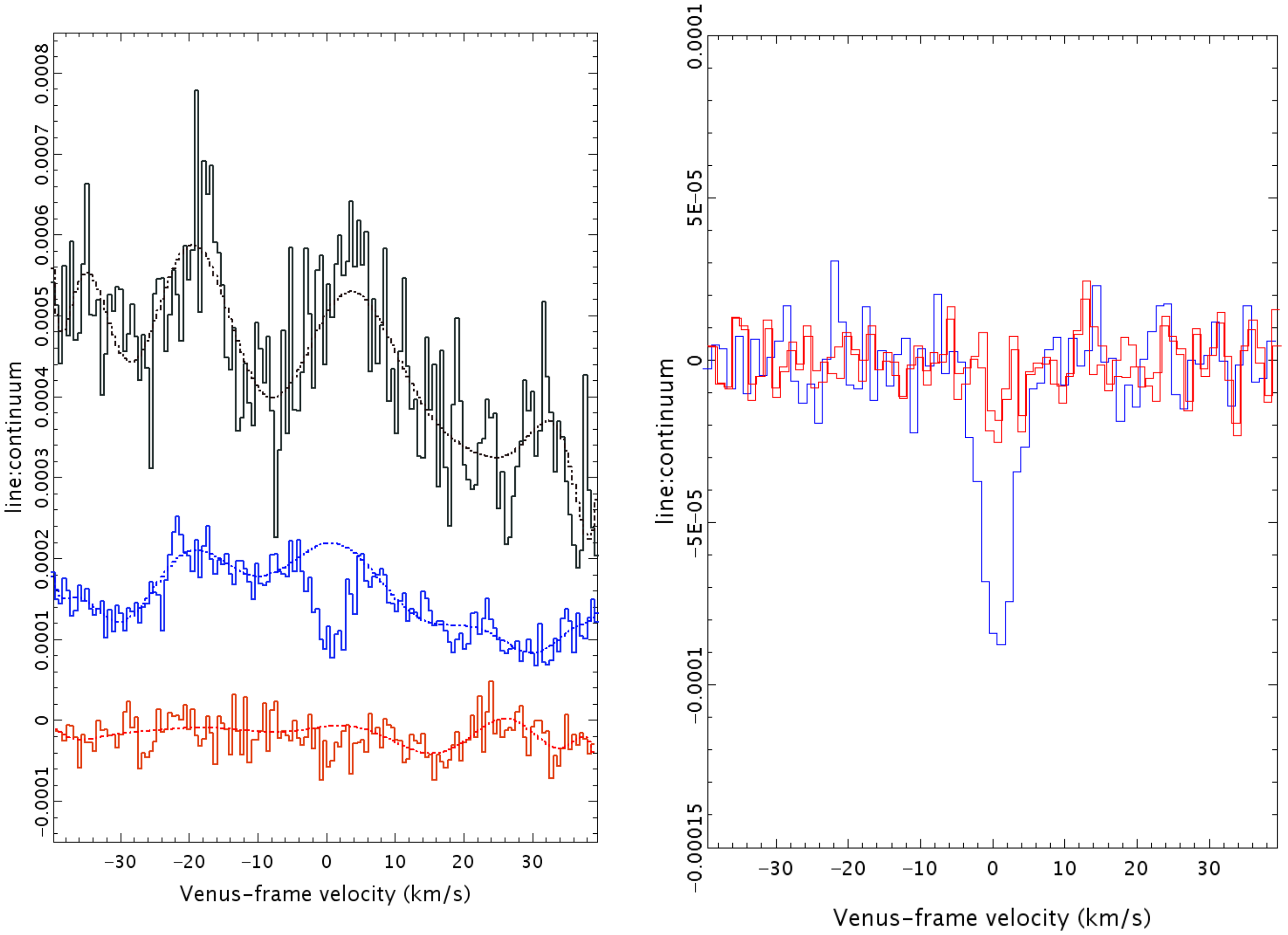

**Supplementary Figure 4.** Instrumental effects in the ALMA data are illustrated. Left panel: $PH_3$ spectra (offset vertically for clarity) illustrating the $12^{th}$-order polynomial functions selected empirically for fitting the spectral ripple seen with ALMA (leading to the results of Figure 2). The planetary zones are polar (top), mid-latitude (centre) and equatorial (bottom) as defined in Table 1. The complexity of the ripple drove the choice of $|v| = 5$ km/s, i.e. line wings more than 5 km/s from the velocity of Venus were forced to zero. This value of $|v|$ was chosen from test-region spectra where the line was clearly visible, and then applied to all the latitudinal bands. The polar spectrum is more noisy because it includes a smaller area (Supplementary Figure 2) and because ripple effects are larger at the planetary limb. Right panel: spectra (red histograms) produced for the whole planet after applying the same reduction procedures to regions of the passband offset by 400 spectral channels either side of the expected line location. This produces narrow artefacts spanning only ~2 spectral channels, much less broad than the real line (blue histogram). The l:c values (integrated over ±5 km/s) of the artefacts are 18 ± 4 % of the value for the real line.

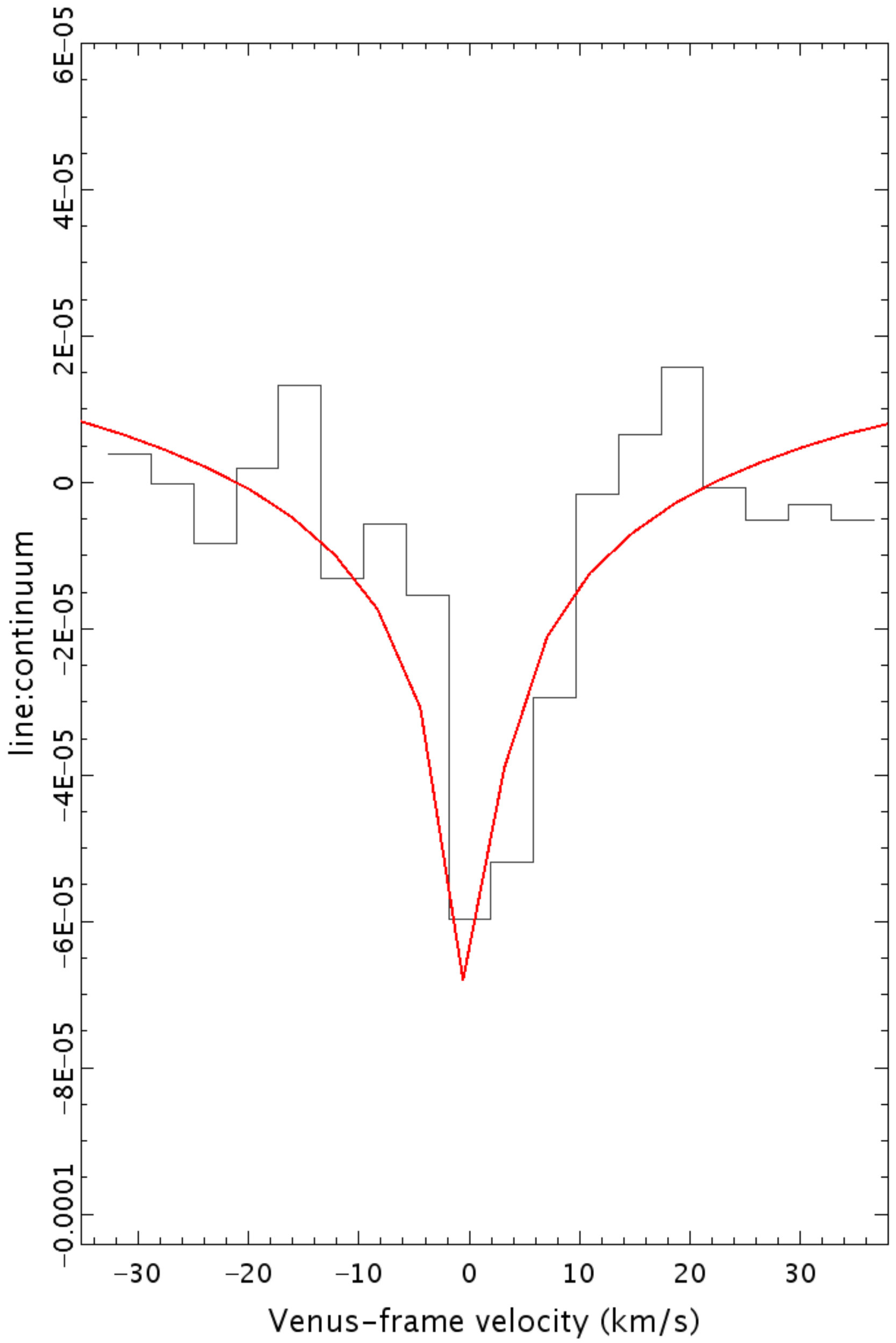

**Supplementary Figure 5.** Deuterated water (HDO) is detected on Venus in the ALMA data. A preliminary reduction of the whole-planet spectrum in the HDO $2_{2,0}$-$3_{1,3}$ transition is shown. The overlaid red curve is from our radiative transfer model, calculated for 2.5 ppb abundance and processed with a 1$^{st}$-order polynomial fit, as for the data. No correction has been made for line-dilution, so the abundance can be significantly under-estimated, depending on the scale over which the molecules are distributed.

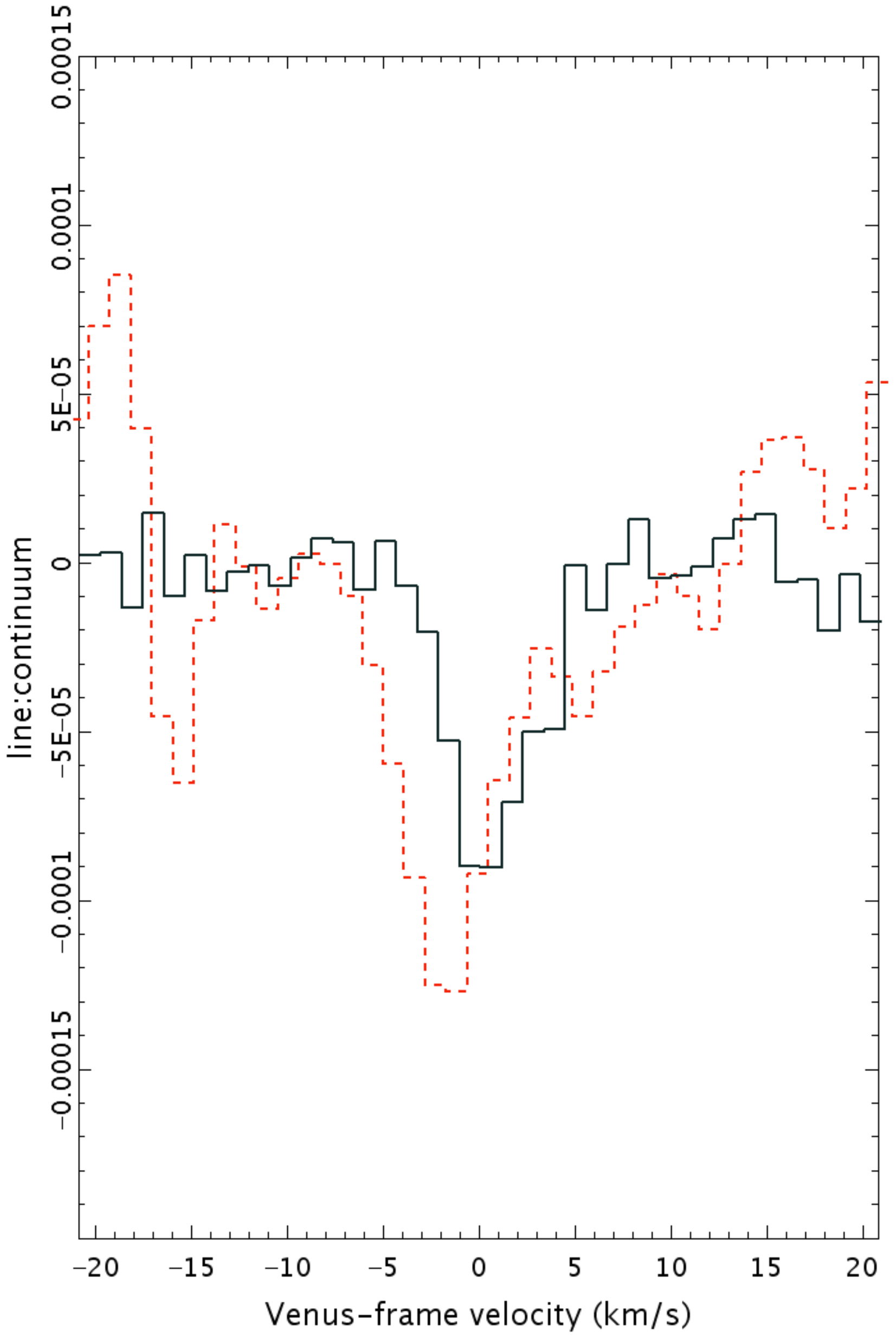

**Supplementary Figure 6.** The whole-planet $PH_3$ 1-0 spectrum (black histogram) from narrowband ALMA data is superposed on the equivalent data recorded simultaneously in the wideband spectral configuration (red dashed histogram). The wideband spectrum has had a $1^{st}$-order polynomial subtracted to correct for mean level and overall slope. The narrowband spectrum (Figure 2) is shown here at the 1.1 km/s resolution of the wideband data. The wideband absorption feature is substantially noisier due to a greater degree of spectral ripple (see e.g. the structure around -15 to -20 km/s), but it supports our $PH_3$ detection, i.e. this detection cannot be attributed to an artefact of one correlator configuration.

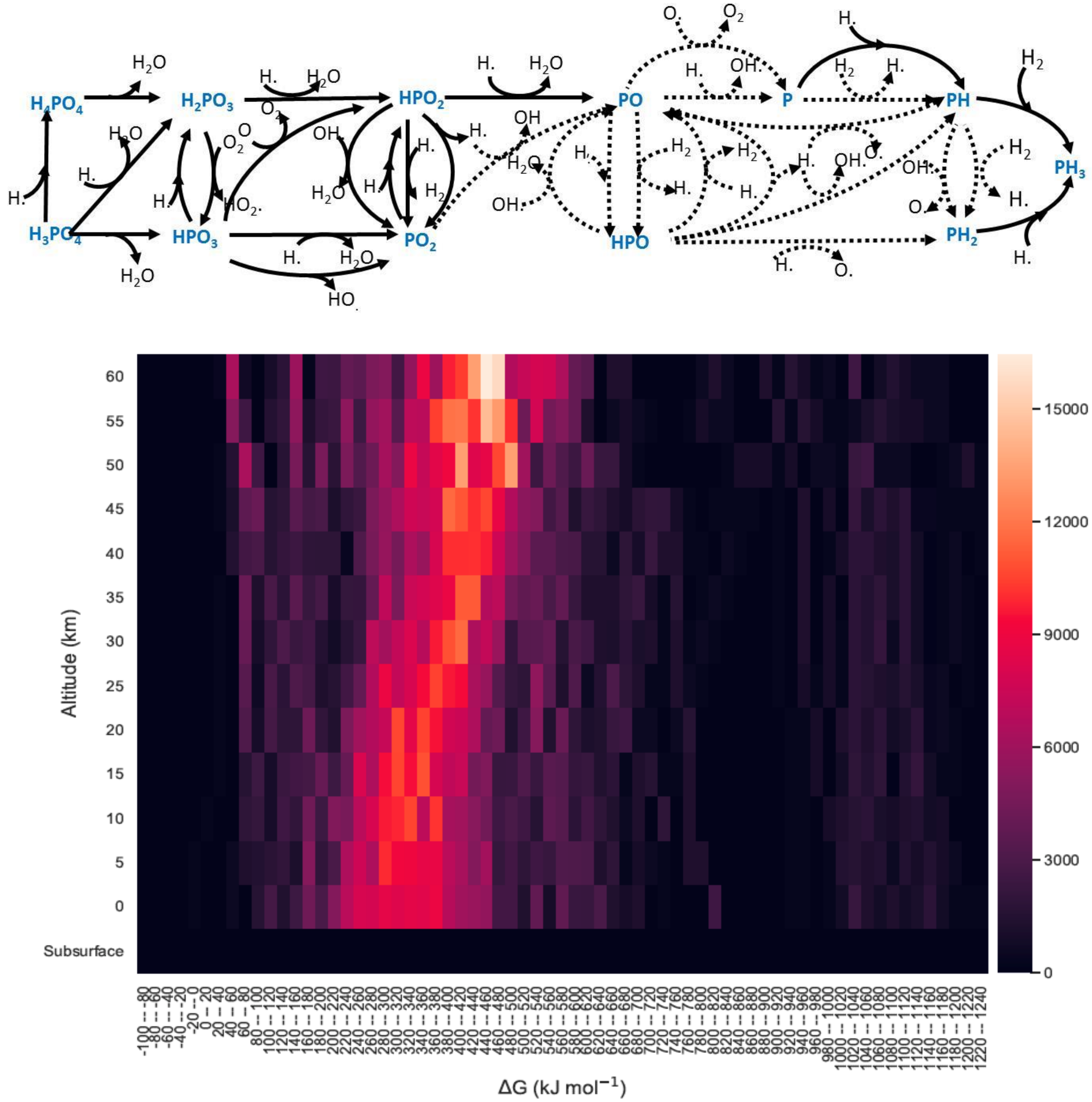

**Supplementary Figure 7.** Chemical pathways and energies are illustrated. Upper panel: Reaction network used to predict maximum possible photochemical production rate for phosphine. Continuous lines are reactions for which kinetic data for the phosphorus species is known. Dotted lines are reactions for which kinetic data for the analogous nitrogen species is known, and was used here. Phosphorus species are shown in blue, reacting radicals in black. Lower panel: Heat map showing that phosphine production is not thermodynamically favored. The plot shows how many reaction/condition combinations there are with given Gibbs free energy as a function of altitude. Y-axis is height above the surface (altitude, in km); columns are bins of data in X, the Gibbs Free Energy (ΔG: -100 to +1240 kJ mol$^{-1}$; 20 kJ mol$^{-1}$ bins). Brighter-colored cells indicate more reactions for a given range of ΔG. There are no reactions occurring in the range where processes would be energetically favorable, i.e. there are no reactions/conditions where ΔG is negative and energy is released.

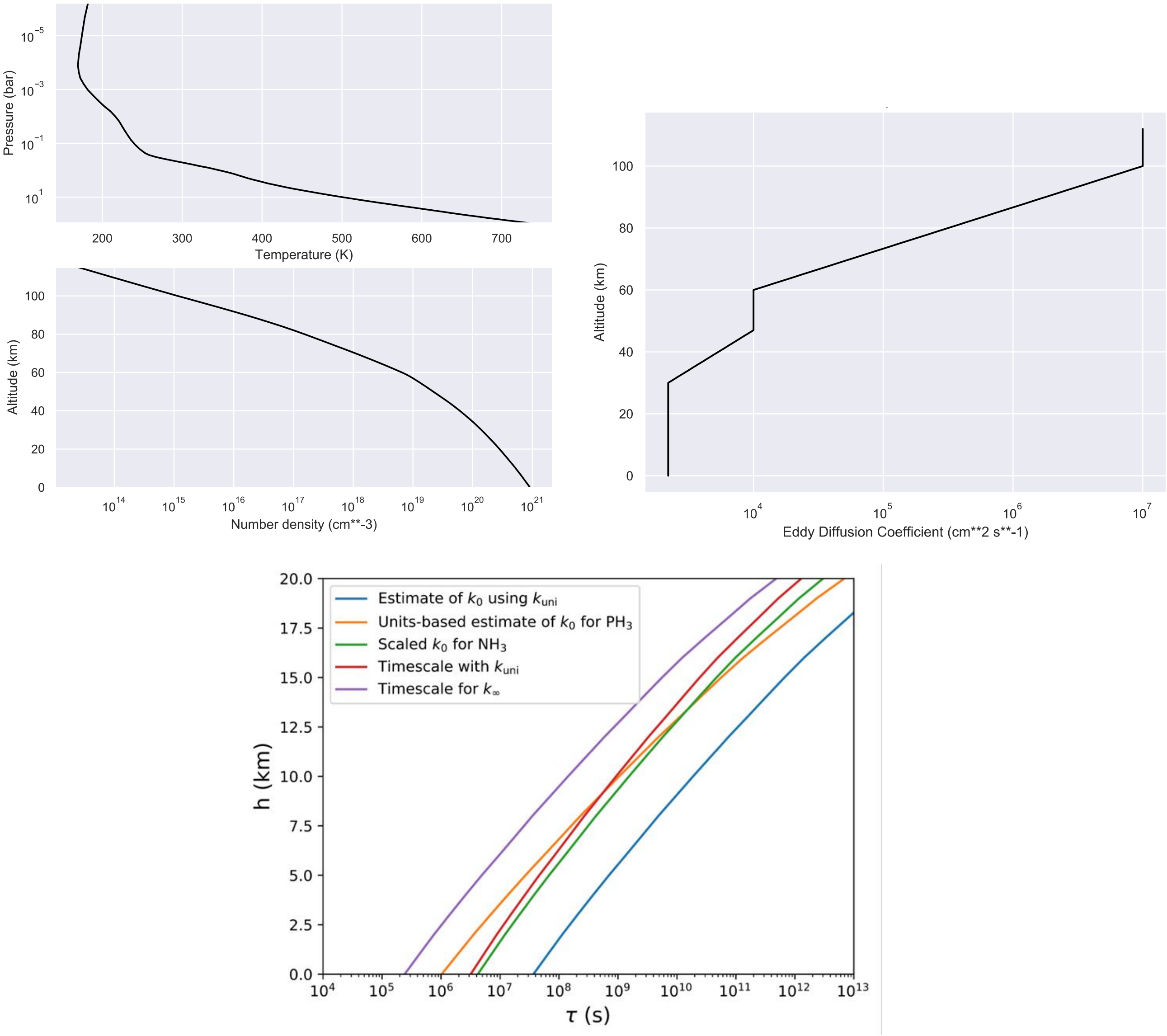

**Supplementary Figure 8.** Inputs to the photochemical modelling are illustrated. Top-left panel: Temperature-pressure profile used in photochemical modelling of the Venusian atmosphere, following refs. 47,57. Top-right panel: Eddy diffusion profile used here in photochemical modelling of the Venusian atmosphere, following refs. 47,57. Lower panel: Decomposition timescale for $PH_3$ as a function of height, derived from the Lindemann approximation of the rate constant, employing a theoretical value of $k_\infty$ ($s^{-1}$) and an approximation of $k_0$ ($cm^3\ s^{-1}$) using $k_{uni}$ ($s^{-1}$, blue), a simple unit-conversion estimate of $k_0$ (orange), a scaled estimate of $k_0$ based on ammonia decomposition (green), the timescale using only $k_{uni}$ (red), and the timescale at the high pressure limit (violet).

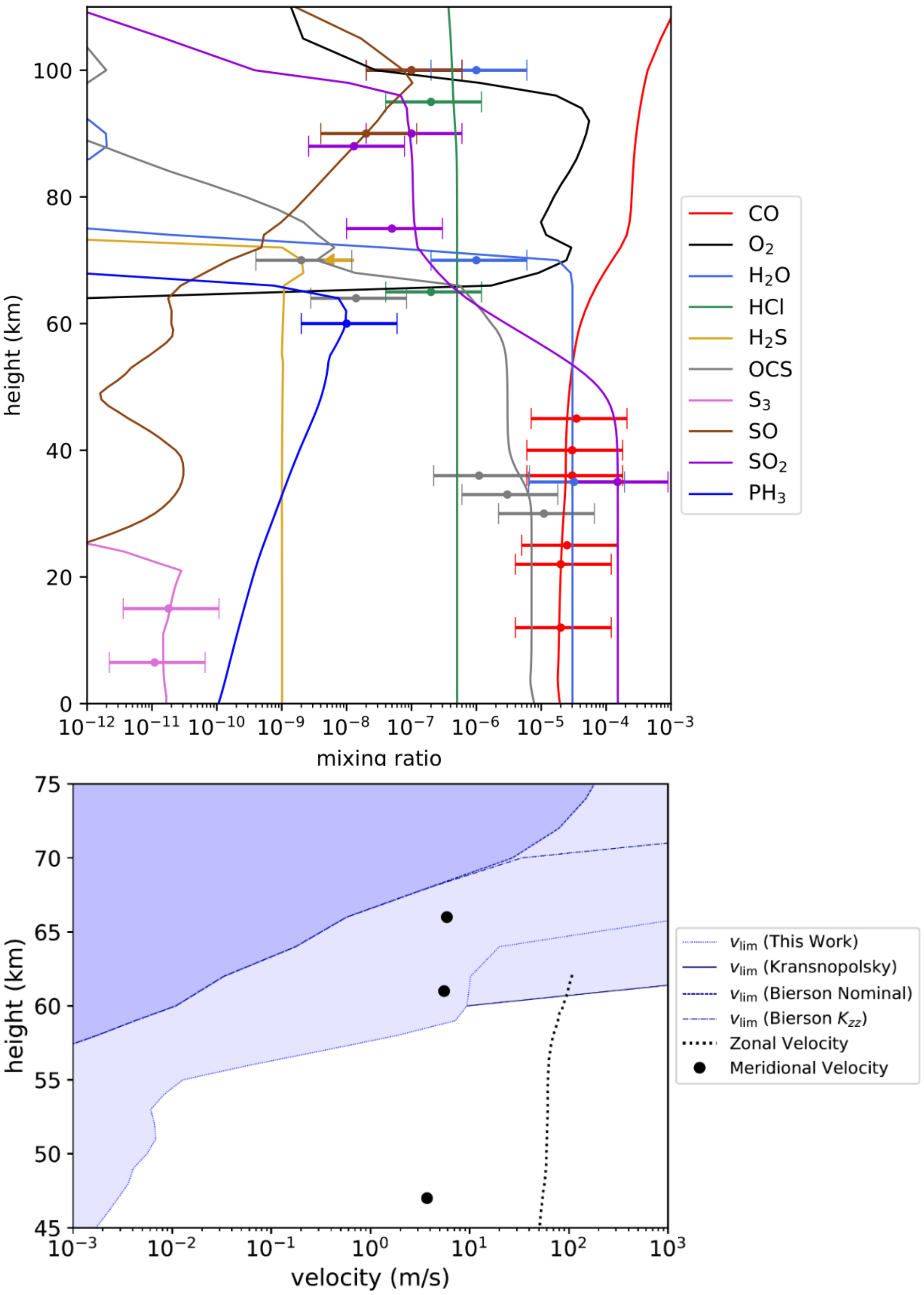

**Supplementary Figure 9.** Context of the photochemical modelling is illustrated. Top panel: Comparison of Venusian-atmosphere model to observations. Mixing ratios of various species are shown versus atmospheric height (km). Error bars span one order of magnitude, to help in comparing model predictions to observations. Bottom panel: Wind velocities that explain observed latitudinal variation, compared to observationally constrained zonal and meridional velocities. Velocities (m/s) are plotted versus atmospheric height (km). The blue shaded regions, bounded by blue lines, show the threshold velocities for the models in question. If the zonal wind velocity exceeds this threshold, then no longitudinal variation is expected, and if the meridional wind velocity exceeds this threshold, no latitudinal variation is expected. Estimated zonal[63] (dashed line) and meridional[64] (circles) wind velocities are also given.

**Supplementary Figure 10**. An overview is presented of the potential pathways for phosphine production in the Venusian environment. None of the known processes can be responsible for the amount of phosphine detected in the Venusian atmosphere.

| **Potential $PH_3$ Production Pathway on Venus** | **Quantitative Barriers for Production Pathway** | **Method** |
|---|---|---|
| Equilibrium thermodynamics of chemical reactions between chemical species in the atmosphere and on the surface | Chemical reactions in Venusian environment are on average 100 kJ/mol too energetically costly (10 - 400 kJ/mol) to proceed spontaneously | Calculation of free energy from known or modeled gas concentrations |
| Equilibrium thermodynamics of chemical reactions in the subsurface | Oxygen fugacity of plausible crust and mantle rocks 8 - 15 orders of magnitude too high to support reduction of phosphate | Calculation of subsurface oxygen fugacity ($fO_2$) |
| Photochemical production by photochemically-generated reactive species | The required forward reaction rates are too low by factors of $10^4 – 10^6$ | UV production of radicals followed by forward kinetic modelling from known and estimated reaction rates |
| Production by lightning | Limited frequency of lightning and low abundance of both atmospheric P species and reducing gases. Less than ppt of $PH_3$ is produced. $PH_3$ production is ~7 orders of magnitude too low to explain detected amounts | Calculations of the maximal efficiency of formation of $PH_3$ upon complete atomization of atmospheric and cloud components containing phosphorus. Literature review of lab experiments on the efficiency of formation of $PH_3$ by lightning discharges. |
| Meteoritic delivery as a source of phosphides and phosphine | The estimated maximal yearly meteoritic delivery of phosphine is ~8 orders of magnitude too low to explain detected amounts | Calculation of the maximum possible amounts of reduced P species delivered assuming their 100% conversion to $PH_3$ |
| Large-scale comet/asteroid impact | Radar mapping of the surface of Venus shows no evidence of a recent large impact | |
| Other endergonic processes as potential sources of phosphine | Solar wind protons and large tribochemical processes cannot be responsible | |

**Supplementary Table 1.** Upper section: list of molecular transitions near the $PH_3$ 1-0 line (highlighted in green), from the Cologne Database for Molecular Spectroscopy (CDMS). The columns list the rest frequencies and their uncertainties, the energy of the lower level of the transition, the transition's quantum numbers, and the chemical species. This selection shows all transitions within 2.5 MHz of the $PH_3$ 1-0 frequency – any transition outside this range would yield a Doppler shift differing from the $PH_3$ 1-0 line by >5σ (for σ ~0.5 km/s: quadratic combination of uncertainties for the ALMA whole-planet spectrum, Table 1). Telluric absorption (at frequencies offset by the Venus-Earth Doppler-shifts) can be ruled out because of much greater angular extent than scales JCMT and ALMA are sensitive to, and telluric line widths exceeding ~100 km/s. The only listed molecule that is plausible for Venus is $SO_2$ (transition highlighted in yellow) and the potential contamination-levels are illustrated in Figure 4. Lower section: list of molecules that we considered as possible contaminants for the $PH_3$ 1-0 line (highlighted in green). Abundances are listed for Venusian molecules that have been measured in-situ or modelled[23]. UNK denotes molecules not been previously measured or predicted on Venus; the estimates in brackets are based on Earth's isotopic ratio. The list covers molecules spectrally active within 80 MHz of the $PH_3$ 1-0 frequency, with listed quantities including transition wavenumber (in GHz, in vacuum), strength (in $cm^{-1}$ / (molecule·$cm^{-2}$), at T=296 K), and rotational quantum numbers (J). The entry in yellow highlights the most likely contaminant, $SO_2$. All other molecules considered are not expected to be present in Venus at sufficiently high concentrations for their listed weak transitions to be likely absorption candidates ('unlikely contributors'). Other species present or possible but with no spectral activity in the passband are collated in the final two rows. Spectral data are from the HITRAN[24], ExoMol[25] and GEISA databases, and from ref. 26.

| freq. (MHz) | error (MHz) | $E_{lower}$ (K) | upper level | lower level | species |
|---|---|---|---|---|---|
| 266942.087 | 0.03 | 26.3 | 7 2 5 | 7 1 7 | CHD2CN |
| 266942.422 | 0.199 | 418.6 | 39 23 16 | 38 23 15 | Glycine, conf. I |
| 266942.422 | 0.199 | 418.6 | 39 23 17 | 38 23 16 | Glycine, conf. I |
| 266942.483 | 2.449 | 206.2 | 33 12 22 | 33 8 25 | i-C3H7CN-13C1 |
| 266942.540 | 16.21 | 912.3 | 65 20 45 | 64 21 44 | s-trans-Propenoic acid |
| 266942.666 | 0.012 | 709.4 | 39 27 13 0 0 | 39 25 15 0 0 | CH3COOH, vt=0 |
| 266942.719 | 0.120 | 61.9 | 8 5 3 | 7 6 2 | c-CC-13-H4O |
| 266943.075 | 0.007 | 1132.7 | 82 64 18 | 81 65 17 | 2-CAB, v=0 |
| 266943.075 | 0.007 | 1132.7 | 82 64 19 | 81 65 16 | 2-CAB, v=0 |
| 266943.325 | 0.0012 | 613.1 | 30 9 21 | 31 8 24 | SO2,v=0 |
| 266943.407 | 0.009 | 1450.4 | 29 0 0 | 28 0 0 | HC3N, v7=4/v5=v7=1 |
| 266943.525 | 0.012 | 513.0 | 29 920 2 0 | 2920 9 1 0 | CH3COOH, Dvt<>0 |
| 266943.536 | 0.086 | 1095.4 | 56 16 40 2 | 55 17 39 1 | CH3OC-13-HO, vt=0,1 |
| 266943.547 | 0.067 | 1133.3 | 48 22 27 2 0 | 47 30 17 1 0 | CH3COOH, Dvt<>0 |
| 266944.038 | 0.013 | 998.3 | 58 16 43 1 | 57 17 40 1 | s-Propanal, v=0 |
| 266944.457 | 0.004 | 376.9 | 46 12 35 0 | 46 10 37 1 | a'GG'g-1,3-Propanediol |
| 266944.510 | 0.443 | 1785.6 | A0 25 76 0 | A0 23 77 1 | gGG'g-1,3-Propanediol |
| 266944.514 | 0.0002 | 0.0 | 1 0 | 0 0 | PH3 |
| 266945.085 | 0.153 | 604.1 | 80 18 62 | 81 17 65 | AA-n-C4H9CN |
| 266945.085 | 0.153 | 604.1 | 80 18 63 | 81 17 64 | AA-n-C4H9CN |
| 266945.102 | 0.003 | 192.8 | 30 3 27 | 29 3 26 | CH2D(oop)CH2CN |
| 266945.315 | 0.005 | 888.5 | 52 17 35 | 53 16 38 | CH3C-13-H2CN, v=0 |
| 266945.315 | 0.005 | 888.5 | 52 17 36 | 53 16 37 | CH3C-13-H2CN, v=0 |
| 266945.781 | 0.057 | 809.4 | 70 33 37 | 70 32 38 | OC(CN)2 |
| 266945.781 | 0.057 | 809.4 | 70 33 38 | 70 32 39 | OC(CN)2 |
| 266946.021 | 0.012 | 998.3 | 58 16 43 0 | 57 17 40 0 | s-Propanal, v=0 |
| 266946.398 | 0.02 | 809.4 | 38 17 22 2 2 | 37 18 19 2 1 | CH3COOH, vt=2 |
| 266946.678 | 4.977 | 1519.2 | 88 9 79 | 88 8 80 | g-n-C3H7CN, v28=1 |
| 266946.960 | 0.0023 | 123.9 | 28 5 23 | 28 2 26 | cis-S2O2 |

| species | type | abundances on Venus (estimates and/or altitude-dependent) | freq. (MHz) | strength ($cm^{-1}$ / $mole.cm^{-2}$)) | quanta ($J_{upper}$-$J_{lower}$) | additional lines | Notes |
|---|---|---|---|---|---|---|---|
| $PH_3$ | molecule | UNK | 266944.5 | 1.27E-22 | 1-0 | | target line |
| $^{32}S^{16}O_2$ | molecule | ~50 ppb – ~150 ppm (Table S3) | 266943.3<br>267006.8 | 3.18E-23<br>4.56E-24 | 30-31<br>15-16 | | modelled contaminant |
| $^{16}O^{16}O^{17}O$ | isotope | UNK (ppb/ppt) | 266925.2 | 5.21E-27 | 23-23 | +15 weaker lines in range | unlikely contributor |
| $^{16}O^{17}O^{16}O$ | isotope | UNK (ppt) | 266932.7 | 5.21E-27 | 23-23 | +12 weaker lines in range | unlikely contributor |
| $^{32}S^{16}O_3$ | molecule | UNK (ppb) | 267024.6<br>266970.6 | 4.91E-28<br>5.82E-28 | 60-59<br>47-48 | | unlikely contributor |
| $^{14}N^{16}O_2$ | molecule | UNK | 266926.4 | 5.08E-27 | 32.5-33.5 | | unlikely contributor |
| $NH_3$ | molecule | UNK | 266992.0 | 1.40E-26 | 3-3 | | unlikely contributor |
| $H^{14}N^{16}O_3$ | molecule | UNK | 266974.5 | 8.95E-25 | 29-29 | +95 weaker lines in range | unlikely contributor |
| $H^{15}N^{16}O_3$ | isotope | UNK | 266952.9 | 4.62E-27 | 14-15 | +2 weaker lines in range | unlikely contributor |
| $^{12}CH_3{}^{16}OH$ | molecule | UNK | 266871.9 | 6.83E-24 | 14-15 | +2 weaker lines in range | unlikely contributor |
| $^{15}N^{14}N^{16}O$ | isotope | UNK | ~266995 (only 4 sig. fig. listed) | 2.46E-25 | 11-10 | | unlikely contributor |
| $CO_2$, $N_2$, $H_2O$, HDO, CO, $H_2SO_4$, $DHSO_4$, OCS, $H_2S$, $O_3$, HCl, DCl, SO, HF, DF, NO | | | | | | | known on Venus, but no spectral activity in passband |
| $ClCO_2$, $S_2O$ (+ isotopologues), $ClSO_2$, ClOS, ClCN, HSD, HNSO, HOCl, DOCl, ClNO, NSCl, $ClO_2$, DNO, DCN | | | | | | | compounds of elements present on Venus; no spectral activity near passband or no data on millimetre transitions |

**Supplementary Table 2.** Initial surface conditions adopted for the atmospheric chemistry.

| **Species** | **Mixing Ratio** |
|---|---|
| $CO_2$ | 0.96 |
| $N_2$ | 0.03 |
| $SO_2$ | $1.5 \times 10^{-4}$ |
| $H_2O$ | $3.0 \times 10^{-5}$ |
| CO | $2.0 \times 10^{-5}$ |
| OCS | $5.0 \times 10^{-6}$ |
| $S_2$ | $7.5 \times 10^{-7}$ |
| HCl | $5.0 \times 10^{-7}$ |
| $S_n$ ($3 \leq n \leq 8$) | $3.3 \times 10^{-7}$ |
| NO | $5.5 \times 10^{-9}$ |
| $H_2$ | $3.0 \times 10^{-9}$ |
| $H_2S$ | $1.0 \times 10^{-9}$ |
| SO | $3.0 \times 10^{-11}$ |
| $ClSO_2$ | $3.0 \times 10^{-11}$ |
| $SO_2Cl_2$ | $1.0 \times 10^{-11}$ |
| HS | $8.0 \times 10^{-13}$ |
| SNO | $1.0 \times 10^{-13}$ |
| SCl | $6.7 \times 10^{-15}$ |
| HSCl | $2.8 \times 10^{-15}$ |
| $Cl_2$ | $1.0 \times 10^{-16}$ |
| S | $7.5 \times 10^{-17}$ |
| H | $7.3 \times 10^{-19}$ |
| OH | $7.3 \times 10^{-19}$ |

**Supplementary Table 3.** Observational constraints on atmospheric concentrations

| Species | Atmospheric Height | Mixing Ratio | Reference |
| --- | --- | --- | --- |
| CO | 12 km | $2 \times 10^{-5}$ | [49] |
| | 22 km | $2 \times 10^{-5}$ | [49] |
| | 25 km | $2.5 \times 10^{-5}$ | [49] |
| | 36 km | $3 \times 10^{-5}$ | [96] |
| | 40 km | $3 \times 10^{-5}$ | [49] |
| | 45 km | $3.5 \times 10^{-5}$ | [49] |
| OCS | 30 km | $1.1 \times 10^{-5}$ | [49] |
| | 33 km | $3 \times 10^{-6}$ | [96] |
| | 36 km | $1.1 \times 10^{-6}$ | [49] |
| | 64 km | $1.4 \times 10^{-8}$ | [97] |
| | 70 km | $2 \times 10^{-9}$ | [97] |
| $SO_2$ | 35 km | $1.5 \times 10^{-4}$ | [96] |
| | 75 km | $5 \times 10^{-8}$ | [98], Average from several observations |
| | 90 km | $1 \times 10^{-7}$ | [98], Average from several observations |
| | 100 km | $1 \times 10^{-7}$ | [98], Average from several observations |
| $H_2O$ | 35 km | $3.2 \times 10^{-5}$ | [96] |
| | 70 km – 100 km | $1 \times 10^{-6}$ | [99], Constant between these heights |
| $H_2S$ | 70 km | $< 2.3 \times 10^{-8}$ | [97], Upper Limit |
| HCl | 65 km – 95 km | $2 \times 10^{-7}$ | [99], Constant between these heights |
| $S_3$ | 6.5 km | $1.1 \times 10^{-11}$ | [49], Heights are approximate |
| | 15 km | $1.8 \times 10^{-11}$ | [49], Heights are approximate |
| SO | 90 km | $2 \times 10^{-8}$ | [98], Average from several observations |
| | 100 km | $1 \times 10^{-7}$ | [98], Average from several observations |
| $PH_3$ | 60 km | $1 \times 10^{-8}$ | This work |

**Supplementary Software 1. (separate file)**

The file is a sequence of commands in a linux shell script that process the JCMT spectra obtainable from the public archive. Reference name of this script is base4_filter9_poly_vshift.sh.

**Supplementary Software 2. (separate file)**

The file is a python script used for initial calibration to produce the ALMA data-cubes we analysed. Reference name is uid___A002_Xd90607_X10526.ms.scriptForCalibration33.py.

**Supplementary Software 3. (separate file)**

The file is a python script used for initial calibration to produce the ALMA data-cubes we analysed. Reference name is uid___A002_Xd90607_X10f75.ms.scriptForCalibration33.py.

**Supplementary Software 4. (separate file)**

The file is a python script used in imaging the ALMA data-cubes. Reference name of the script is Venus_imaging.py.

**Supplementary References:**

63 Counselman, C. C., Gourevitch, S. A., King, R. W., Loriot, G. B. & Ginsberg, E. S. Zonal and meridional circulation of the lower atmosphere of Venus determined by radio interferometry. *Journal of Geophysical Research: Space Physics* **85**, 8026-8030 (1980).

64 Sánchez-Lavega, A. *et al.* Variable winds on Venus mapped in three dimensions. *Geophysical Research Letters* **35** (2008).

65 Cardelino, B. H. *et al.* Semiclassical calculation of reaction rate constants for homolytical dissociation reactions of interest in organometallic vapor-phase epitaxy (OMVPE). *The Journal of Physical Chemistry A* **107**, 3708-3718 (2003).

66 Davidson, D. F., Kohse-Höinghaus, K., Chang, A. Y. & Hanson, R. K. A pyrolysis mechanism for ammonia. *International journal of chemical kinetics* **22**, 513-535 (1990).

67 Kaye, J. A. & Strobel, D. F. Phosphine photochemistry in the atmosphere of Saturn. *Icarus* **59**, 314-335 (1984).

68 Bolshova, T. A. & Korobeinichev, O. P. Promotion and inhibition of a hydrogen—oxygen flame by the addition of trimethyl phosphate. *Combustion, Explosion and Shock Waves* **42**, 493-502 (2006).

69 Rock, P. A. *Chemical thermodynamics: principles and applications*. (The Macmillan Company, 1969).

70 Greiner, W., Neise, L. & Stöcker, H. *Thermodynamics and statistical mechanics*. (Springer Science & Business Media, 2012).

71 Perrot, P. *A to Z of Thermodynamics*. (Oxford University Press on Demand, 1998).

72 Andreichikov, B. M. in *Lunar and Planetary Science Conference.*

73 Marcq, E., Mills, F. P., Parkinson, C. D. & Vandaele, A. C. Composition and chemistry of the neutral atmosphere of Venus. *Space Science Reviews* **214**, 10 (2018).

74 Oyama, V. *et al.* Pioneer Venus gas chromatography of the lower atmosphere of Venus. *Journal of Geophysical Research: Space Physics* **85**, 7891-7902 (1980).

75 Taylor, F. W. & Hunten, D. M. in *Encyclopedia of the Solar System* p305-322 (Elsevier, 2014).
76 Vandaele, A. C. *et al.* Sulfur dioxide in the Venus atmosphere: I. Vertical distribution and variability. *Icarus* **295**, 16-33 (2017).
77 Smrekar, S. E., Stofan, E. R. & Mueller, N. in *Encyclopedia of the Solar System* 323-341 (Elsevier, 2014).
78 Taylor, F. W., Svedhem, H. & Head, J. W. Venus: the atmosphere, climate, surface, interior and near-space environment of an Earth-like planet. *Space Science Reviews* **214**, 35 (2018).
79 Holland, H. D. *The chemical evolution of the atmosphere and oceans*. (Princeton University Press, 1984).
80 Glindemann, D., de Graaf, R. M. & Schwartz, A. W. Chemical Reduction of Phosphate on the Primitive Earth. *Origins of life and evolution of the biosphere* **29**, 555-561, doi:10.1023/A:1006622900660 (1999).
81 Glindemann, D., Edwards, M. & Schrems, O. Phosphine and methylphosphine production by simulated lightning—a study for the volatile phosphorus cycle and cloud formation in the earth atmosphere. *Atmospheric Environment* **38**, 6867-6874, doi:https://doi.org/10.1016/j.atmosenv.2004.09.002 (2004).
82 Peucker-Ehrenbrink, B. Accretion of extraterrestrial matter during the last 80 million years and its effect on the marine osmium isotope record. *Geochimica et Cosmochimica Acta* **60**, 3187-3196, doi:https://doi.org/10.1016/0016-7037(96)00161-5 (1996).
83 Emiliani, C. *Planet Earth: Cosmology, Geology, and the Evolution of Life and Environment*. (Cambridge University Press, 1992).
84 Geist, V., Wagner, G., Nolze, G. & Moretzki, O. Investigations of the meteoritic mineral (Fe,Ni)3P. *Crystal Research and Technology* **40**, 52-64, doi:doi:10.1002/crat.200410307 (2005).
85 Cockell, C. S. Life on Venus. *Planetary and Space Science* **47**, 1487-1501 (1999).
86 Grinspoon, D. H. *Venus revealed: a new look below the clouds of our mysterious twin planet* (1997).
87 Schulze-Makuch, D., Grinspoon, D. H., Abbas, O., Irwin, L. N. & Bullock, M. A. A sulfur-based survival strategy for putative phototrophic life in the Venusian atmosphere. *Astrobiology* **4**, 11-18 (2004).
88 Wu, Z. *et al.* The near-UV absorber OSSO and its isomers. *Chemical Communications* **54**, 4517-4520, doi:10.1039/C8CC00999F (2018).
89 Bains, W., Petkowski, J. J., Sousa-Silva, C. & Seager, S. Trivalent Phosphorus and Phosphines as Components of Biochemistry in Anoxic Environments. *Astrobiology* **19**, 885-902, doi:10.1089/ast.2018.1958 (2019).
90 Way, M. J. & Del Genio, A. in *EPSC-DPS Joint Meeting* Vol. 13 (Geneva, Switzerland, 2019).
91 Way, M. J. *et al.* Was Venus the first habitable world of our solar system? *Geophysical research letters* **43**, 8376-8383 (2016).
92 Phillips, R. J. *et al.* Impact craters and Venus resurfacing history. *Journal of Geophysical Research: Planets* **97**, 15923-15948 (1992).
93 Gillmann, C. & Tackley, P. Atmosphere/mantle coupling and feedbacks on Venus. *Journal of Geophysical Research: Planets* **119**, 1189-1217 (2014).

94 Amato, P. *et al.* Metatranscriptomic exploration of microbial functioning in clouds. *Scientific reports* **9** (2019).
95 Bryan, N. C., Christner, B. C., Guzik, T. G., Granger, D. J. & Stewart, M. F. Abundance and survival of microbial aerosols in the troposphere and stratosphere. *The ISME journal*, 1-11 (2019).
96 Marcq, E. *et al.* A latitudinal survey of CO, OCS, H2O, and SO2 in the lower atmosphere of Venus: Spectroscopic studies using VIRTIS-H. *Journal of Geophysical Research: Planets* **113** (2008).
97 Krasnopolsky, V. A. High-resolution spectroscopy of Venus: Detection of OCS, upper limit to H2S, and latitudinal variations of CO and HF in the upper cloud layer. *Icarus* **197**, 377-385 (2008).
98 Belyaev, D. A. *et al.* Vertical profiling of SO2 and SO above Venus' clouds by SPICAV/SOIR solar occultations. *Icarus* **217**, 740-751 (2012).
99 Bertaux, J.L. *et al.* A warm layer in Venus' cryosphere and high-altitude measurements of HF, HCl, H2O and HDO. *Nature* **450**, 646-649 (2007).